
\input phyzzx



\message{For printing the figures, you need to have the
files psfig.tex, 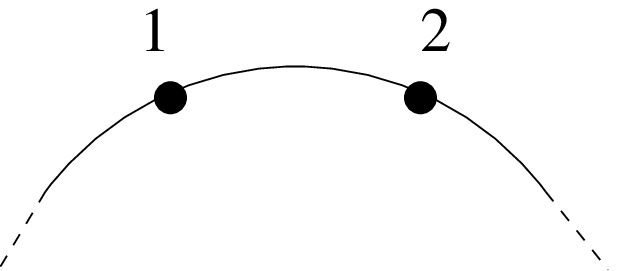 and chain.eps
in your area. Do you have these files? Enter y/n :    }
\read -1 to \figcount
\if y\figcount \message{This will come out with the figures}
\else \message{This will come out without the figures}\fi

\if y\figcount
    \input psfig
\else\fi

\catcode`\@=12        
%
\newbox\hdbox%
\newcount\hdrows%
\newcount\multispancount%
\newcount\ncase%
\newcount\ncols
\newcount\nrows%
\newcount\nspan%
\newcount\ntemp%
\newdimen\hdsize%
\newdimen\newhdsize%
\newdimen\parasize%
\newdimen\spreadwidth%
\newdimen\thicksize%
\newdimen\thicksz
\newdimen\thinsize%
\newdimen\tablewidth%
\newif\ifcentertables%
\newif\ifendsize%
\newif\iffirstrow%
\newif\iftableinfo%
\newtoks\dbt%
\newtoks\hdtks%
\newtoks\savetks%
\newtoks\tableLETtokens%
\newtoks\tabletokens%
\newtoks\widthspec%
%
%
%
%
\tableinfotrue%
\catcode`\@=11
%
%
\def\tstrut{\vrule height3.1ex depth1.2ex width0pt}%
\def\and{\char`\&}
\def\tablerule{\noalign{\hrule height\thinsize depth0pt}}%
\thicksize=1.5pt
\thinsize=0.6pt
\thicksz=1.5pt
\def\thickrule{\noalign{\hrule height\thicksize depth0pt}}%
\def\ctr#1{\hfil\ #1\hfil}%
%
%
%
%
\tablewidth=-\maxdimen%
\spreadwidth=-\maxdimen%
\def\tabskipglue{0pt plus 1fil minus 1fil}%
%
%
\centertablestrue%
%
%
%
%
\parasize=4in%
\gdef\ARGS{########}
\gdef\headerARGS{####}
\def\@mpersand{&}
{\catcode`\|=13
\gdef\letbarzero{\let|0}
\gdef\letbartab{\def|{&&}}%
\gdef\letvbbar{\let\vb|}%
}
{\catcode`\&=4
\def\ampskip{&\omit\hfil&}
\catcode`\&=13
\let&0
\xdef\letampskip{\def&{\ampskip}}%
\gdef\letnovbamp{\let\novb&\let\tab&}
}
\def\begintable{
   \begingroup%
   \catcode`\|=13\letbartab\letvbbar%
   \catcode`\&=13\letampskip\letnovbamp%
   \def\multispan##1{
      \omit \mscount##1%
      \multiply\mscount\tw@\advance\mscount\m@ne%
      \loop\ifnum\mscount>\@ne \sp@n\repeat%
   }
   \def\|{%
      &\omit\widevline&%
   }%
   \ruledtable
}
\long\def\ruledtable#1\endtable{%
%
%
%
   \offinterlineskip
   \tabskip 0pt
   \def\widevline{\vrule width\thicksz}
   \def\endrow{\@mpersand\omit\hfil\crnorm\@mpersand}%
   \def\crthick{\@mpersand\crnorm\thickrule\@mpersand}%
   \def\crthickneg##1{\@mpersand\crnorm\thickrule
	  \noalign{{\skip0=##1\vskip-\skip0}}\@mpersand}%
   \def\crnorule{\@mpersand\crnorm\@mpersand}%
   \def\crnoruleneg##1{\@mpersand\crnorm
	  \noalign{{\skip0=##1\vskip-\skip0}}\@mpersand}%
   \let\nr=\crnorule
   \def\endtable{\@mpersand\crnorm\thickrule}%
   \let\crnorm=\cr
%
%
   \edef\cr{\@mpersand\crnorm\tablerule\@mpersand}%
   \def\crneg##1{\@mpersand\crnorm\tablerule
	  \noalign{{\skip0=##1\vskip-\skip0}}\@mpersand}%
   \let\ctneg=\crthickneg
   \let\nrneg=\crnoruleneg
   \the\tableLETtokens
%
%
   \tabletokens={&#1}
%
%
   \countROWS\tabletokens\into\nrows%
   \countCOLS\tabletokens\into\ncols%
%
%
   \advance\ncols by -1%
   \divide\ncols by 2%
   \advance\nrows by 1%
%
%
   \iftableinfo %
      \immediate\write16{[Nrows=\the\nrows, Ncols=\the\ncols]}%
   \fi%
%
%
   \ifcentertables
      \ifhmode \par\fi
      \line{
      \hss
   \else %
      \hbox{%
   \fi
      \vbox{%
	 \makePREAMBLE{\the\ncols}
	 \edef\next{\preamble}
	 \let\preamble=\next
	 \makeTABLE{\preamble}{\tabletokens}
      }
      \ifcentertables \hss}\else }\fi
   \endgroup
   \tablewidth=-\maxdimen
   \spreadwidth=-\maxdimen
}
\def\makeTABLE#1#2{
   {
   \let\ifmath0
   \let\header0
   \let\multispan0
%
%
   \ncase=0%
   \ifdim\tablewidth>-\maxdimen \ncase=1\fi%
   \ifdim\spreadwidth>-\maxdimen \ncase=2\fi%
   \relax
%
   \ifcase\ncase %
      \widthspec={}%
   \or %
      \widthspec=\expandafter{\expandafter t\expandafter o%
		 \the\tablewidth}%
   \else %
      \widthspec=\expandafter{\expandafter s\expandafter p\expandafter r%
		 \expandafter e\expandafter a\expandafter d%
		 \the\spreadwidth}%
   \fi %
   \xdef\next{
      \halign\the\widthspec{%
      #1
      \noalign{\hrule height\thicksize depth0pt}
%
      \the#2\endtable
%
      }
   }
   }
   \next
}
\def\makePREAMBLE#1{
   \ncols=#1
   \begingroup
   \let\ARGS=0
   \edef\xtp{\widevline\ARGS\tabskip\tabskipglue%
   &\ctr{\ARGS}\tstrut}
   \advance\ncols by -1
   \loop
      \ifnum\ncols>0 %
      \advance\ncols by -1%
      \edef\xtp{\xtp&\vrule width\thinsize\ARGS&\ctr{\ARGS}}%
   \repeat
   \xdef\preamble{\xtp&\widevline\ARGS\tabskip0pt%
   \crnorm}
   \endgroup
}
\def\countROWS#1\into#2{
   \let\countREGISTER=#2%
   \countREGISTER=0%
   \expandafter\ROWcount\the#1\endcount%
}%
\def\ROWcount{%
   \afterassignment\subROWcount\let\next= %
}%
\def\subROWcount{%
   \ifx\next\endcount %
      \let\next=\relax%
   \else%
      \ncase=0%
      \ifx\next\cr %
	 \global\advance\countREGISTER by 1%
	 \ncase=0%
      \fi%
      \ifx\next\endrow %
	 \global\advance\countREGISTER by 1%
	 \ncase=0%
      \fi%
      \ifx\next\crthick %
	 \global\advance\countREGISTER by 1%
	 \ncase=0%
      \fi%
      \ifx\next\crnorule %
	 \global\advance\countREGISTER by 1%
	 \ncase=0%
      \fi%
      \ifx\next\crthickneg %
	 \global\advance\countREGISTER by 1%
	 \ncase=0%
      \fi%
      \ifx\next\crnoruleneg %
	 \global\advance\countREGISTER by 1%
	 \ncase=0%
      \fi%
      \ifx\next\crneg %
	 \global\advance\countREGISTER by 1%
	 \ncase=0%
      \fi%
      \ifx\next\header %
	 \ncase=1%
      \fi%
      \relax%
      \ifcase\ncase %
	 \let\next\ROWcount%
      \or %
	 \let\next\argROWskip%
      \else %
      \fi%
   \fi%
   \next%
}
\def\counthdROWS#1\into#2{%
\dvr{10}%
   \let\countREGISTER=#2%
   \countREGISTER=0%
\dvr{11}%
\dvr{13}%
   \expandafter\hdROWcount\the#1\endcount%
\dvr{12}%
}%
\def\hdROWcount{%
   \afterassignment\subhdROWcount\let\next= %
}%
\def\subhdROWcount{%
   \ifx\next\endcount %
      \let\next=\relax%
   \else%
      \ncase=0%
      \ifx\next\cr %
	 \global\advance\countREGISTER by 1%
	 \ncase=0%
      \fi%
      \ifx\next\endrow %
	 \global\advance\countREGISTER by 1%
	 \ncase=0%
      \fi%
      \ifx\next\crthick %
	 \global\advance\countREGISTER by 1%
	 \ncase=0%
      \fi%
      \ifx\next\crnorule %
	 \global\advance\countREGISTER by 1%
	 \ncase=0%
      \fi%
      \ifx\next\header %
	 \ncase=1%
      \fi%
\relax%
      \ifcase\ncase %
	 \let\next\hdROWcount%
      \or%
	 \let\next\arghdROWskip%
      \else %
      \fi%
   \fi%
   \next%
}%
{\catcode`\|=13\letbartab
\gdef\countCOLS#1\into#2{%
   \let\countREGISTER=#2%
   \global\countREGISTER=0%
   \global\multispancount=0%
   \global\firstrowtrue
   \expandafter\COLcount\the#1\endcount%
   \global\advance\countREGISTER by 3%
   \global\advance\countREGISTER by -\multispancount
}%
\gdef\COLcount{%
   \afterassignment\subCOLcount\let\next= %
}%
{\catcode`\&=13%
\gdef\subCOLcount{%
   \ifx\next\endcount %
      \let\next=\relax%
   \else%
      \ncase=0%
      \iffirstrow
	 \ifx\next& %
	    \global\advance\countREGISTER by 2%
	    \ncase=0%
	 \fi%
	 \ifx\next\span %
	    \global\advance\countREGISTER by 1%
	    \ncase=0%
	 \fi%
	 \ifx\next| %
	    \global\advance\countREGISTER by 2%
	    \ncase=0%
	 \fi
	 \ifx\next\|
	    \global\advance\countREGISTER by 2%
	    \ncase=0%
	 \fi
	 \ifx\next\multispan
	    \ncase=1%
	    \global\advance\multispancount by 1%
	 \fi
	 \ifx\next\header
	    \ncase=2%
	 \fi
	 \ifx\next\cr	    \global\firstrowfalse \fi
	 \ifx\next\endrow   \global\firstrowfalse \fi
	 \ifx\next\crthick  \global\firstrowfalse \fi
	 \ifx\next\crnorule \global\firstrowfalse \fi
	 \ifx\next\crnoruleneg \global\firstrowfalse \fi
	 \ifx\next\crthickneg  \global\firstrowfalse \fi
	 \ifx\next\crneg       \global\firstrowfalse \fi
      \fi
\relax
      \ifcase\ncase %
	 \let\next\COLcount%
      \or %
	 \let\next\spancount%
      \or %
	 \let\next\argCOLskip%
      \else %
      \fi %
   \fi%
   \next%
}%
\gdef\argROWskip#1{%
   \let\next\ROWcount \next%
}
\gdef\arghdROWskip#1{%
   \let\next\ROWcount \next%
}
\gdef\argCOLskip#1{%
   \let\next\COLcount \next%
}
}
}
\def\spancount#1{
   \nspan=#1\multiply\nspan by 2\advance\nspan by -1%
   \global\advance \countREGISTER by \nspan
   \let\next\COLcount \next}%
\def\dvr#1{\relax}%
\def\header#1{%
\dvr{1}{\let\cr=\@mpersand%
\hdtks={#1}%
\counthdROWS\hdtks\into\hdrows%
\advance\hdrows by 1%
\ifnum\hdrows=0 \hdrows=1 \fi%
\dvr{5}\makehdPREAMBLE{\the\hdrows}%
\dvr{6}\getHDdimen{#1}%
{\parindent=0pt\hsize=\hdsize{\let\ifmath0%
\xdef\next{\valign{\headerpreamble #1\crnorm}}}\dvr{7}\next\dvr{8}%
}%
}\dvr{2}}
\def\makehdPREAMBLE#1{
\dvr{3}%
\hdrows=#1
{
\let\headerARGS=0%
\let\cr=\crnorm%
\edef\xtp{\vfil\hfil\hbox{\headerARGS}\hfil\vfil}%
\advance\hdrows by -1
\loop
\ifnum\hdrows>0%
\advance\hdrows by -1%
\edef\xtp{\xtp&\vfil\hfil\hbox{\headerARGS}\hfil\vfil}%
\repeat%
\xdef\headerpreamble{\xtp\crcr}%
}
\dvr{4}}
\def\getHDdimen#1{%
\hdsize=0pt%
\getsize#1\cr\end\cr%
}
\def\getsize#1\cr{%
\endsizefalse\savetks={#1}%
\expandafter\lookend\the\savetks\cr%
\relax \ifendsize \let\next\relax \else%
\setbox\hdbox=\hbox{#1}\newhdsize=1.0\wd\hdbox%
\ifdim\newhdsize>\hdsize \hdsize=\newhdsize \fi%
\let\next\getsize \fi%
\next%
}%
\def\lookend{\afterassignment\sublookend\let\looknext= }%
\def\sublookend{\relax%
\ifx\looknext\cr %
\let\looknext\relax \else %
   \relax
   \ifx\looknext\end \global\endsizetrue \fi%
   \let\looknext=\lookend%
    \fi \looknext%
}%
%
%
\def\tablelet#1{%
   \tableLETtokens=\expandafter{\the\tableLETtokens #1}%
}%
\catcode`\@=12
%


\font\eighti=cmmi8                          \skewchar\eighti='177
\font\eightsy=cmsy8                          \skewchar\eightsy='60
\font\eightsl=cmsl8
\font\eightit=cmti8
\def\noblackbox{\overfullrule=0pt}
\noblackbox
\def\bold#1{\setbox0=\hbox{$#1$}%
     \kern-.025em\copy0\kern-\wd0
     \kern.05em\copy0\kern-\wd0
     \kern-.025em\raise.0433em\box0 }
\def\unlock{\catcode`@=11} 
\def\lock{\catcode`@=12} 
\def\Buildrel#1\under#2{\mathrel{\mathop{#2}\limits_{#1}}}
\def\llongrarrow{\hbox to 40pt{\rightarrowfill}}

%
 \newtoks\slashfraction
 \slashfraction={.13}
 \def\slash#1{\setbox0\hbox{$ #1 $}
 \setbox0\hbox to \the\slashfraction\wd0{\hss \box0}/\box0 }
 \unlock
 \def\leftrightarrowfill{$\m@th\mathord-\mkern-6mu%
   \cleaders\hbox{$\mkern-2mu\mathord-\mkern-2mu$}\hfill
   \mkern-6mu\mathord\leftrightarrow$}
 \def\overlrarrow#1{\vbox{\ialign{##\crcr
       \leftrightarrowfill\crcr\noalign{\kern-\p@\nointerlineskip}
       $\hfil\displaystyle{#1}\hfil$\crcr}}}
 \lock
%

{\obeyspaces\global\let =\ }   
%
\def\papersize{       \hsize=35pc\vsize=50pc\hoffset=1cm\voffset=1.3cm
              \pagebottomfiller=0pc
              \skip\footins=\bigskipamount\normalspace}
\def\lettersize{\hsize=6.5in\vsize=8.5in\hoffset=0cm\voffset=1.6cm
              \pagebottomfiller=\letterbottomskip
              \skip\footins=\smallskipamount
              \multiply\skip\footins by 3
              \singlespace}
\papers
%
\catcode`\@=11 
\newif\ifletterstyle                
\letterstylefalse             
\def\letters{\lettersize\letterstyletrue
   \headline=\letterheadline \footline=\letterfootline
   \immediate\openout\labelswrite=\jobname.lab}
\def\iftpub{\afterassignment\iftp@b\toks@}
\def\iftp@b{\edef\n@xt{\Pubnum={UFIFT--HEP--\the\toks@}}\n@xt}
\let\pubnum=\iftpub
\expandafter\ifx\csname eightrm\endcsname\relax
    \let\eightrm=\ninerm \let\eightbf=\ninebf \fi
\catcode`\@=12 
%
   
 

\unlock
\def\eightpoint{\relax
    \textfont0=\eightrm          \scriptfont0=\eightrm
    \scriptscriptfont0=\fiverm
    \def\rm{\fam0 \eightrm \f@ntkey=0 }\relax
    \textfont1=\eighti           \scriptfont1=\eighti
    \scriptscriptfont1=\fivei
    \def\oldstyle{\fam1 \eighti \f@ntkey=1 }\relax
    \textfont2=\eightsy          \scriptfont2=\eightsy
    \scriptscriptfont2=\fivesy
    \textfont3=\tenex          \scriptfont3=\tenex
    \scriptscriptfont3=\tenex
    \def\it{\fam\itfam \eightit \f@ntkey=4 }\textfont\itfam=\eightit
    \def\sl{\fam\slfam \eightsl \f@ntkey=5 }\textfont\slfam=\eightsl
    \def\bf{\fam\bffam \eightbf \f@ntkey=6 }\textfont\bffam=\eightbf
        \scriptfont\bffam=\eightbf     \scriptscriptfont\bffam=\fivebf
    \def\tt{\fam\ttfam \eighttt \f@ntkey=7 }\textfont\ttfam=\eighttt
    \setbox\strutbox=\hbox{\vrule height 4pt depth 3pt width\z@}
    \samef@nt}
\lock
\def\boxit#1{\vbox{\hrule\hbox{\vrule\kern3pt
             \vbox{\kern3pt#1\kern3pt}\kern3pt\vrule}\hrule}}
 \newdimen\str

\def\fboxit#1#2{\vbox{\hrule height #1 \hbox{\vrule width #1
           \kern3pt \vbox{\kern3pt#2\kern3pt}\kern3pt \vrule width #1 }
           \hrule height #1 }}

\def\fillbox#1{\hbox to #1{\vbox to #1{\vfil}\hfil}}
\def\dotbox#1{\hbox to #1{\vbox to 10pt{\vfil}\hss $\cdots$ \hss}}
\def\ggenbox#1#2{\vbox to 10pt{\vss \hbox to #1{\hss #2  \hss} \vss}}


\catcode`\@=11 
\newtoks\foottokens
\let\labelfont=\Tenpoint      
\def\MakeFromBox{\gl@bal\setbox\FromLabelBox=\vbox{\labelfont
     \ialign{##\hfil\cr \the\sendername \the\FromAddress \crcr }}}
\def\smallsize{\relax
\def\eightpoint{\relax
\textfont0=\eightrm  \scriptfont0=\sixrm
\scriptscriptfont0=\fiverm
\def\rm{\fam0 \eightrm \f@ntkey=0}\relax
\textfont1=\eighti  \scriptfont1=\sixi
\scriptscriptfont1=\fivei
\def\oldstyle{\fam1 \eighti \f@ntkey=1}\relax
\textfont2=\eightsy  \scriptfont2=\sixsy
\scriptscriptfont2=\fivesy
\textfont3=\tenex  \scriptfont3=\tenex
\scriptscriptfont3=\tenex
    \def\it{\fam\itfam \eightit \f@ntkey=4 }\textfont\itfam=\eightit
\def\sl{\fam\slfam \eightsl \f@ntkey=5 }\textfont\slfam=\eightsl
\def\bf{\fam\bffam \eightbf \f@ntkey=6 }\textfont\bffam=\eightbf
\scriptfont\bffam=\sixbf   \scriptscriptfont\bffam=\sixbf
\def\tt{\fam\ttfam \eighttt \f@ntkey=7 }
\def\caps{\fam\cpfam \tencp \f@ntkey=8 }\textfont\cpfam=\tencp
\setbox\strutbox=\hbox{\vrule height 7.35pt depth 3.02pt width\z@}
\samef@nt}
\normalbaselineskip = 16.60pt plus 0.166pt minus 0.083pt
\normallineskip = 1.25pt plus 0.08pt minus 0.08pt
\normallineskiplimit = 1.25pt
\normaldisplayskip = 16.60pt plus 4.15pt minus 8.3pt
\normaldispshortskip = 4.98pt plus 3.32pt
\normalparskip = 4.98pt plus 1.67pt minus .83pt
\skipregister = 4.15pt plus 1.67pt minus 1.25pt
\def\Eightpoint{\eightpoint \relax
  \ifsingl@\subspaces@t2:5;\else\subspaces@t3:5;\fi
  \ifdoubl@ \multiply\baselineskip by 5
            \divide\baselineskip by 4\fi }
\parindent=16.67pt
\itemsize=25pt
\thinmuskip=2.5mu
\medmuskip=3.33mu plus 1.67mu minus 3.33mu
    \thickmuskip=4.17mu plus 4.17mu
\def\thinspace{\kern .13889em }
\def\negthinspace{\kern-.13889em }
\def\enspace{\kern.416667em }
\def\enskip{\hskip.416667em\relax}
\def\quad{\hskip.83333em\relax}
\def\qquad{\hskip1.66667em\relax}
\def\crr{\cropen{8.3333pt}}
\labelwidth=4.5in
\let\labelfont=\Eightpoint
\let\letterhead=\FLOHEAD      
\def\Vfootnote##1{\insert\footins\bgroup
   \interlinepenalty=\interfootnotelinepenalty \floatingpenalty=20000
   \singl@true\doubl@false\Eightpoint
   \splittopskip=\ht\strutbox \boxmaxdepth=\dp\strutbox
   \leftskip=\footindent \rightskip=\z@skip
   \parindent=0.5\footindent \parfillskip=0pt plus 1fil
   \spaceskip=\z@skip \xspaceskip=\z@skip \footnotespecial
   \Textindent{##1}\footstrut\futurelet\next\fo@t}%
\def\attach##1{\step@ver{\strut^{\mkern 1.6667mu ##1} } }
\def\inserttable ##1##2##3%
    {%
    \tbldef {##1}{##3}\goodbreak%
        \midinsert
      \smallskip
      \hbox{\singlespace \hskip 0.5cm
              \vtop{\parshape=2 0cm 10.8cm 1.3cm 9.5cm
                    \noindent{\bf\Table{##1}}.\enspace ##3}
              \hfil}
      ##2
      \smallskip
    \endinsert
    }
\def\sure{y}
\def\insertfigureold ##1##2##3%
    {%
    \figdef {##1}{##3}\goodbreak%
    \midinsert
      \smallskip
      ##2
      \hbox{\singlespace\hskip 0.5cm
              \vtop{\parshape=2 0cm 10.8cm
                      1.6cm 9.2cm \noindent{\bf\Figure{##1}}.
                      \enspace ##3}
              \hfil}
        \smallskip
    \endinsert
    }%
\def\references{\par\penalty-300\vskip\chapterskip
        \spacecheck\chapterminspace
      \line{\twelverm\hfil REFERENCES\hfil}
      \nobreak\vskip\headskip\penalty 30000
      \reflist{}}
\def\figures{\par\penalty-300\vskip\chapterskip
        \spacecheck\chapterminspace
      \line{\twelverm\hfil FIGURE CAPTIONS\hfil}
      \nobreak\vskip\headskip\penalty 30000
      \figlist{}}
\def\tables{\par\penalty-300\vskip\chapterskip
        \spacecheck\chapterminspace
      \line{\twelverm\hfil TABLE CAPTIONS\hfil}
      \nobreak\vskip\headskip\penalty 30000
      \tbllist{}}
\def\PH@SR@V{\doubl@true\baselineskip=20.08pt plus .1667pt minus .0833pt
             \parskip = 2.5pt plus 1.6667pt minus .8333pt }
    \def\author##1{\vskip\frontpageskip\titlestyle{\tencp ##1}\nobreak}
\def\address##1{\par\kern 4.16667pt\titlestyle{\tenpoint\it ##1}}
\def\andaddress{\par\kern 4.16667pt \centerline{\sl and} \address}
\def\UFL{\address{Department of Physics\break
      University of Florida, Gainesville, FL 32611}}
\def\abstract{\vskip\frontpageskip\centerline{\twelverm ABSTRACT}
              \vskip\headskip }
\def\submit##1{\par\nobreak\vfil\nobreak\medskip
   \centerline{Submitted to \sl ##1}}
\def\doeack{\foot{Work supported in part by the Department of Energy
              under grant  DE--FG05--86ER--40272.}}
\def\nsfack{\foot{Work supported by National Science Foundation
              Grant  PHY 84--16030A01.}}
\def\cases##1{\left\{\,\vcenter{\Tenpoint\m@th
    \ialign{$####\hfil$&\quad####\hfil\crcr##1\crcr}}\right.}
\def\matrix##1{\,\vcenter{\Tenpoint\m@th
    \ialign{\hfil$####$\hfil&&\quad\hfil$####$\hfil\crcr
      \mathstrut\crcr\noalign{\kern-\baselineskip}
     ##1\crcr\mathstrut\crcr\noalign{\kern-\baselineskip}}}\,}
\Tenpoint
    }
\newdimen\fullhsize
\newbox\leftcolumn
\def\twoinone{
\smallsize
\def\papersize{
              \voffset=-.23truein
              \vsize=7truein
              \baselineskip=16pt plus 2pt minus 1pt
              \fullhsize=10truein\hsize=4.75truein
                \hoffset=-.54truein
              \skip\footins=\bigskipamount}
\def\lettersize{\voffset=.31truein
              \vsize=6.38truein
              \baselineskip=16pt plus 2pt minus 1pt
              \fullhsize=10truein\hsize=4.75truein
                \hoffset=-.48truein
              \skip\footins=\smallskipamount
                \multiply\skip\footins by3}
\papers               
    \let\lr=L
\output={\if L\lr
              \global\setbox\leftcolumn=\columnbox \advancepageno
              \global\let\lr=R
       \else  \getitout \advancepageno
              \global\let\lr=L\fi
       \ifnum\outputpenalty>-20000 \else\dosupereject\fi}
}             
      \def\columnbox{\leftline{
              \vbox{\ifletterstyle\makeheadline\fi
                      \pagebody\makefootline}}}
      \def\fullline{\hbox to\fullhsize}
      \def\getitout{\shipout\vbox{\fullline{\box\leftcolumn
              \hfil {\leftline{
              \vbox{\makeheadline
              \pagebody\makefootline}}} }}}
%
\catcode`\@=12 
    %
%
%
\newcount      \ObjClass
\chardef\ClassNum     = 0
\chardef\ClassMisc    = 1
\chardef\ClassEqn     = 2
\chardef\ClassRef     = 3
\chardef\ClassFig     = 4
\chardef\ClassTbl     = 5
\chardef\ClassThm     = 6
\chardef\ClassStyle     = 7
    \chardef\ClassDef       = 8
\edef\NumObj  {\ObjClass = \ClassNum   \relax}
\edef\MiscObj {\ObjClass = \ClassMisc  \relax}
\edef\EqnObj  {\ObjClass = \ClassEqn   \relax}
\edef\RefObj  {\ObjClass = \ClassRef   \relax}
\edef\FigObj  {\ObjClass = \ClassFig   \relax}
\edef\TblObj  {\ObjClass = \ClassTbl   \relax}

\edef\StyleObj  {\ObjClass = \ClassStyle \relax}
\edef\DefObj    {\ObjClass = \ClassDef   \relax}
%
%
\def\gobble    #1{}%
\def\trimspace   #1 \end{#1}%
\def\ifundefined #1{\expandafter \ifx \csname#1\endcsname \relax}%
\def\trimprefix  #1_#2\end{\expandafter \string \csname #2\endcsname}%
\def\skipspace #1#2#3\end%
    {%
    \def \temp {#2}%
    \ifx \temp\space \skipspace #1#3\end
    \else \gdef #1{#2#3}\fi
        }%
\def\stylename#1{\expandafter\expandafter\expandafter
    \gobble\expandafter\string\the#1}
\ifundefined {protect} \let\protect=\relax \fi
\catcode`\@=11
\let\rel@x=\relax
\def\relaxtest{\rel@x}
\catcode`\@=12
\def\checkchapterlabel%
    {%
        {\protect\if\chapterlabel\relaxtest
      \global\let\chapterlabel=\relax\fi}
    }%
\begingroup
\catcode`\<=1 \catcode`\{=12
\catcode`\>=2 \catcode`\}=12
\xdef\LBrace<{>%
\xdef\RBrace<}>%
\endgroup
%
%
\newcount\equanumber \equanumber=0
    \newcount\eqnumber   \eqnumber=0
\newif\ifleftnumbers \leftnumbersfalse

\def\(#1)%
     {%
        \ifnum \equanumber<0 \eqnumber=-\equanumber
          \advance\eqnumber by -1 \else
            \eqnumber = \equanumber\fi
        \ifmmode\ifinner(\eqnum{#1})\else
        \ifleftnumbers\leqno(\eqnum{#1})\ifdraft{\rm[#1]}\fi
            \else\eqno(\eqnum{#1})\ifdraft{\rm[#1]}\fi\fi\fi
      \else(\eqnum{#1})\fi\ifnum%
          \equanumber<0 \global\equanumber=-\eqnumber\global\advance
            \equanumber by -1\else\global\equanumber=\eqnumber\fi
     }%
\def\mideq(#1)%
     {%
      \ifleftnumbers \leqinsert{$\(#1)$} \else
      \eqinsert{$\(#1)$} \fi
     }%
\def\eqnum #1%
    {%
    \LookUp Eq_#1 \using\eqnumber\neweqnum
    {\rm\label}%
    }%
\def\neweqnum #1#2%
    {%
    \checkchapterlabel
    {\protect\xdef\eqnoprefix{\ifundefined{chapterlabel}
      \else\chapterlabel.\fi}}
    \ifmmode \xdef #1{\eqnoprefix #1}
        \else\message{Undefined equation \string#1 in non-math mode.}%
           \xdef #1{\relax}
           \global\advance \eqnumber by -1
        \fi
    \EqnObj \SaveObject{#1}{#2}
    }%
\everydisplay = {\expandafter \let\csname Eq_\endcsname=\relax
               \expandafter \let\csname Eq_?\endcsname=\relax}%
%
    %
\newcount\tablecount \tablecount=0
\def\Table  #1{Table~\tblnum {#1}}%
\def\tblnum #1{\TblObj \LookUp Tbl_#1 \using\tablecount
      \SaveObject \label\ifdraft [#1]\fi}%
\def\tbldef #1{\TblObj \SaveContents {Tbl_#1}}%
\def\tbllist  {\TblObj \ListObjects}%
%
\def\inserttable #1#2#3%
    {%
    \tbldef {#1}{#3}\goodbreak%
    \midinsert
      \smallskip
      \hbox{\singlespace
            \vtop{\titlestyle{{\Tenpoint{\caps\Table{#1}}\break #3}}}
           }%
      #2
      \smallskip%
    \endinsert
    }%
    \def\topinserttable #1#2#3%
    {%
    \tbldef {#1}{#3}\goodbreak%
    \topinsert
      \smallskip
      \hbox{\singlespace
            \vtop{\titlestyle{{\Tenpoint{\caps\Table{#1}}\break #3}}}
           }%
      #2
      \smallskip%
    \endinsert
    }%
%
%
\newcount\figurecount \figurecount=0
\def\Figure #1{Figure~\fignum {#1}}%
\def\Fig    #1{Fig.~\fignum {#1}}%
\def\fignum #1{\FigObj \LookUp Fig_#1 \using\figurecount
     \SaveObject \label\ifdraft [#1]\fi}%
    \def\figdef #1{\FigObj \SaveContents {Fig_#1}}%
\def\figlist  {\FigObj \ListObjects}%
%
\def\sure{y}
\def\insertfigure #1#2#3
    {%
    \figdef {#1}{#3}%
    \midinsert
      \bigskip
      \if y\figcount
      #2
      \else
         {\smallskip FIGURE UNAVAILABLE
      }\fi
      \smallskip
     \hbox{\singlespace\hskip 0.5cm
              \vtop{\parshape=2 0cm 10.8cm
                      1.6cm 9.2cm \noindent{\bf\Figure{#1}}.
                      \enspace #3}
              \hfil}
      \smallskip%
    \endinsert
    }%
\def\topinsertfigure #1#2#3%
    {%
        \figdef {#1}{#3}%
    \topinsert
      \bigskip
      #2
      \hbox{  \singlespace
              \hskip 0.4in
              \vtop{\parshape=2 0pt 362pt 32pt 330pt
                    \noindent{\Tenpoint{\caps\Fig{#1}}.\enspace #3}}
              \hfil}
      \smallskip%
    \endinsert
    }%
%
%
%
    \newcount\theoremcount \theoremcount=0
%
%
%
%
%
%
%
%
%
%
%
%
%
%
\newcount\referencecount \referencecount=0
\newcount\refsequence \refsequence=0
\newcount\lastrefno   \lastrefno=-1
%
\def\NPrefs{\let\refmark=\NPrefmark \let\refitem=\NPrefitem}

%
\def\refsymbol#1{\refrange#1-\end}%
\def\[#1]#2%
      {%
      \if.#2\rlap.\refmark{\refsymbol{#1}}\let\refendtok=\relax%
      \else\if,#2\rlap,\refmark{\refsymbol{#1}}\let\refendtok=\relax%
      \else\refmark{\refsymbol{#1}}\let\refendtok=#2\fi\fi%
      \discretionary{}{}{}\refendtok}%
\def\refrange #1-#2\end%
    {%
    \refnums #1,\end
    \def \temp {#2}%
    \ifx \temp\empty \else -\expandafter\refrange \temp\end \fi
    }%
\def\refnums #1,#2\end%
    {%
    \def \temp {#1}%
    \ifx \temp\empty \else \skipspace \temp#1\end\fi
    \ifx \temp\empty
      \ifcase \refsequence
          \or\or ,\number\lastrefno
            \else  -\number\lastrefno
      \fi
      \global\lastrefno = -1
      \global\refsequence = 0
    \else
      \RefObj \edef\temp {Ref_\temp\space}%
      \expandafter \LookUp \temp \using\referencecount\SaveObject
      \global\advance \lastrefno by 1
      \edef \temp {\number\lastrefno}%
      \ifx \label\temp
          \global\advance\refsequence by 1
      \else
          \global\advance\lastrefno by -1
          \ifcase \refsequence
              \or ,%
              \or ,\number\lastrefno,%
          \else   -\number\lastrefno,%
          \fi
          \label
          \global\refsequence = 1
          \ifx\suffix\empty
              \global\lastrefno = \label
          \else
                \global\lastrefno = -1
          \fi
      \fi
      \refnums #2,\end
    \fi
    }%
%
%
%
%
%
\def\reflist  {\RefObj \ListObjects}%
\def\Refer #1{Ref.[\refsymbol{#1}]}%
%
%
\newif\ifSaveFile
\newif\ifnotskip
\newwrite\SaveFile
    \let\IfSelect=\iftrue
\edef\savefilename {\jobname.aux}%
\def\Def#1#2%
    {%
    \expandafter\gdef\noexpand#1{#2}%
    \DefObj \SaveObject {#2}{\expandafter\gobble\string#1}%
}%
\def\savestate%
    {%
    \ifundefined {chapternumber} \else
      \NumObj \SaveObject {\number\chapternumber}{chapternumber} \fi
        \ifundefined {appendixnumber} \else
      \NumObj \SaveObject {\number\appendixnumber}{appendixnumber} \fi
    \ifundefined {sectionnumber} \else
      \NumObj \SaveObject {\number\sectionnumber}{sectionnumber} \fi
    \ifundefined {pagenumber} \else
      \advance\pagenumber by 1
      \NumObj \SaveObject {\number\pagenumber}{pagenumber}%
      \advance\pagenumber by -1 \fi
    \NumObj \SaveObject {\number\equanumber}{equanumber}%
    \NumObj \SaveObject {\number\tablecount}{tablecount}%
        \NumObj \SaveObject {\number\figurecount}{figurecount}%
    \NumObj \SaveObject {\number\theoremcount}{theoremcount}%
    \NumObj \SaveObject {\number\referencecount}{referencecount}%
    \checkchapterlabel
    \ifundefined {chapterlabel} \else
      {\protect\xdef\chaplabel{\chapterlabel}}
      \MiscObj \SaveObject \chaplabel {chapterlabel} \fi
    \ifundefined {chapterstyle} \else
      \StyleObj \SaveObject
           {\stylename{\chapterstyle}}{chapterstyle} \fi
    \ifundefined {appendixstyle} \else
      \StyleObj \SaveObject
           {\stylename{\appendixstyle}}{appendixstyle}\fi
}%
\def\Contents #1{\ObjClass=-#1 \SaveContents}%
\def\Define #1#2#3%
    {%
    \ifnum #1=\ClassNum
      \global \csname#2\endcsname = #3 %
    \else \ifnum #1=\ClassStyle
      \global \csname#2\endcsname\expandafter=
        \expandafter{\csname#3\endcsname} %
    \else \ifnum #1=\ClassDef
        \expandafter\gdef\csname#2\endcsname{#3} %
    \else
      \expandafter\xdef \csname#2\endcsname {#3} \fi\fi\fi %
    \ObjClass=#1 \SaveObject {#3}{#2}%
    }%
\def\SaveObject #1#2%
    {%
    \ifSaveFile \else \OpenSaveFile \fi
    \immediate\write\SaveFile
      {%
      \noexpand\IfSelect\noexpand\Define
      {\the\ObjClass}{#2}{#1}\noexpand\fi
      }%
    }%
\def\SaveContents #1%
    {%
    \ifSaveFile \else \OpenSaveFile \fi
    \BreakLine
    \SaveLine {#1}%
    }%
\begingroup
        \catcode`\^^M=\active %
\gdef\BreakLine %
    {%
    \begingroup %
    \catcode`\^^M=\active %
    \newlinechar=`\^^M %
    }%
\gdef\SaveLine #1#2%
    {%
    \toks255={#2}%
    \immediate\write\SaveFile %
      {%
      \noexpand\IfSelect\noexpand\Contents
      {-\the\ObjClass}{#1}\LBrace\the\toks255\RBrace\noexpand\fi%
      }%
    \endgroup %
    }%
\endgroup
\def\ListObjects #1%
    {%
    \ifSaveFile \CloseSaveFile \fi
    \let \IfSelect=\GetContents \ReadFileList #1,\savefilename,\end
       \let \IfSelect=\IfDoObject  \input \savefilename
    \let \IfSelect=\iftrue
    }%
\def\ReadFileList #1,#2\end%
    {%
    \def \temp {#1}%
    \ifx \temp\empty \else \skipspace \temp#1\end \fi
    \ifx \temp\empty \else \input #1 \fi
    \def \temp {#2}%
    \ifx \temp\empty \else \ReadFileList #2\end \fi
    }%
\def\GetContents #1#2#3%
    {%
    \notskipfalse
    \ifnum \ObjClass=-#2
      \expandafter\ifx \csname #3\endcsname
        \relax \else \notskiptrue \fi
    \fi
    \ifnotskip \expandafter \DefContents \csname #3_\endcsname
    }%
\def\DefContents #1#2{\toks255={#2} \xdef #1{\the\toks255}}%
    \def\IfDoObject #1#2%
    {%
    \notskipfalse \ifnum \ObjClass=#2
        \notskiptrue\fi \ifnotskip \DoObject
    }%
\def\DoObject #1#2%
    {%
    \ifnum \ObjClass = \ClassTbl      \par\noindent Table~#2.
    \else \ifnum \ObjClass = \ClassFig        \par\noindent Figure~#2.
    \else \ifnum \ObjClass = \ClassRef  \refitem{#2}
    \else \item {#2.}
    \fi\fi\fi
    \ifdraft\edef\temp
       {\trimprefix #1\end}[\expandafter\gobble \temp]~\fi
    \expandafter\ifx \csname #1_\endcsname \relax
      \ifdraft\relax\else\edef\temp {\trimprefix #1\end}%
      [\expandafter\gobble \temp]\fi%
    \else
      \csname #1_\endcsname
    \fi
    }%
\def\OpenSaveFile   {\immediate\openout\SaveFile=\savefilename
                   \global\SaveFiletrue}%
\def\CloseSaveFile  {\immediate\closeout
                 \SaveFile \global\SaveFilefalse}%
%
%
\def\LookUp #1 #2\using#3#4%
    {%
    \expandafter \ifx\csname#1\endcsname \relax
        \global\advance #3 by 1
        \expandafter \xdef \csname#1\endcsname {\number #3}%
        \let \newlabelfcn=#4%
          \ifx \newlabelfcn\relax \else
           \expandafter \newlabelfcn \csname#1\endcsname {#1}%
         \fi
    \fi
    \xdef \label  {\csname#1\endcsname}%
    \gdef \suffix {#2}%
    \ifx \suffix\empty \else
        \xdef \suffix {\expandafter\trimspace \suffix\end}%
        \xdef \label  {\label\suffix}%
    \fi
    }%
   %
%
%
\newcount\appendixnumber        \appendixnumber=0
\newtoks\appendixstyle                \appendixstyle={\Alphabetic}
\newif\ifappendixlabel                \appendixlabelfalse
\def\APPEND#1{\par\penalty-300\vskip\chapterskip
      \spacecheck\chapterminspace
      \global\chapternumber=\number\appendixnumber
      \global\advance\appendixnumber by 1
      \chapterstyle\expandafter=\expandafter{\the\appendixstyle}
\chapterreset
     \titlestyle{Appendix\ifappendixlabel~\chapterlabel\fi.~ {#1}}
      \nobreak\vskip\headskip\penalty 30000}
%

%
%
%
\def\references#1{\par\penalty-300\vskip\chapterskip\spacecheck
       \chapterminspace\line{\fourteenrm\hfil References\hfil}
       \nobreak\vskip\headskip\penalty 30000\reflist{#1}}
\def\figures#1{\par\penalty-300\vskip\chapterskip\spacecheck
     \chapterminspace\line{\fourteenrm\hfil Figure Captions\hfil}
     \nobreak\vskip\headskip\penalty 30000\figlist{#1}}
\def\tables#1{\par\penalty-300\vskip\chapterskip\spacecheck
      \chapterminspace\line{\fourteenrm\hfil Table Captions\hfil}
       \nobreak\vskip\headskip\penalty 30000\tbllist{#1}}
\newif\ifdraft\draftfalse
\newcount\yearltd\yearltd=\year\advance\yearltd by -1900
\def\draft{\drafttrue
    \def\draftdate{preliminary draft:
        \number\month/\number\day/\number\yearltd\ \ \hourmin}%
        \paperheadline={\hfil\draftdate} \headline=\paperheadline
        {\count255=\time\divide\count255 by 60
                   \xdef\hourmin{\number\count255}
            \multiply\count255 by-60\advance\count255 by\time
    \xdef\hourmin{\hourmin:\ifnum\count255<10 0\fi\the\count255} }
      \message{draft mode}  }
%
\def\slacpub{ \Pubnum={$\caps SLAC - PUB - \the\pubnum $}}
%
%

\def\frac#1/#2{\leavevmode\kern.1em\raise.5ex
              \hbox{\the\scriptfont0
              #1}\kern-.1em/\kern-.15em
              \lower.25ex\hbox{\the\scriptfont0 #2}}
%
%

\def\frac#1#2{{#1 \over #2}}

%

%
\IfSelect \Contents
{-3}{Ref_Beg}{M.A.B. B\'eg and R.C. Furlong,
{\sl Phys. Rev.} {\bf D31} (1985) 1370.}\fi
\IfSelect \Contents
{-3}{Ref_bergman1}{O. Bergman,
{\sl Phys. Rev.} {\bf D46} (1992) 5474.}\fi
\IfSelect \Contents
{-3}{Ref_JackiwPi}{R. Jackiw and S.Y. Pi,
{\sl Phys. Rev.} {\bf D31} (1985) 848.}\fi
\IfSelect \Contents
{-3}{Ref_pert}{E.L. Feinberg,
{\sl Sov. Phys. Usp. (Eng. Trans.} {\bf 5} (1963) 753;
E. Corinaldesi and F. Rafeli,
{\sl Am. J. Phys.} {\bf 46} (1978) 1185;
K.M. Purcell and W.C. Henneberger,
{\sl Am. J. Phys.} {\bf 46} (1978) 1255.}\fi
\IfSelect \Contents
{-3}{Ref_bergman2}{O. Bergman and G. Lozano,
{\sl Ann. Phys. (NY)} {\bf 229} (1994) 416.}\fi
\IfSelect \Contents
{-3}{Ref_MIT}{M. Leblanc, G. Lozano and H. Min,
{\sl Ann. Phys. (NY)} {\bf 219} (1992) 328.}\fi
\IfSelect \Contents
{-3}{Ref_supercs}{C. Lee, K. Lee and E.J. Weinberg,
{\sl Phys. Lett.} {\bf B243} (1990) 105.}\fi
\IfSelect \Contents
{-3}{Ref_thornmosc}{C. B. Thorn,
``Reformulating String Theory with the 1/N Expansion,''
in {\it Sakharov Memorial Lectures in Physics},
Ed. L. V. Keldysh and V. Ya. Fainberg, Nova Science Publishers Inc.,
Commack, New York, 1992. hep-th/9405069}\fi
\IfSelect \Contents
{-3}{Ref_klebanovs}{I. Klebanov and L. Susskind,
{\sl Nucl. Phys.} {\bf B309} (1988) 175.}\fi
\IfSelect \Contents
{-3}{Ref_thooftlargen}{G. 't Hooft,
{\sl Nucl. Phys.} {\bf B72} (1974) 461.}\fi
\IfSelect \Contents
{-3}{Ref_thooftbh}{G. 't Hooft,
{\sl Nucl. Phys.} {\bf B342} (1990) 471;
``On the Quantization of Space and Time,'' {\it Proc. of the
4th Seminar on Quantum Gravity}, 25\dash29 May 1987, Moscow, USSR,
ed. M. A. Markov \etal, (World Scientific Press, 1988);
``Dimensional Reduction in Quantum Gravity,'' Utrecht preprint,
THU-93/26, GR-QC/9310026.}\fi
\IfSelect \Contents
{-3}{Ref_nielsenfishnet}{H. B. Nielsen and P. Olesen,
{\sl Phys. Lett.} {\bf 32B} (1970) 203;
B. Sakita and M. A. Virasoro,
{\sl Phys. Rev. Lett.} {\bf 24} (1970) 1146.}\fi
\IfSelect \Contents
{-3}{Ref_gilest}{R. Giles and C. B. Thorn,
{\sl Phys. Rev.} {\bf D16} (1977) 366.}\fi
\IfSelect \Contents
{-3}{Ref_thornfishnet}{C. B. Thorn,
{\sl Phys. Rev.} {\bf D17} (1978) 1073.}\fi
\IfSelect \Contents
{-3}{Ref_bardakcis}{K. Bardakci and S. Samuel,
{\sl Phys. Rev.} {\bf D16} (1977) 2500.}\fi
\IfSelect \Contents
{-3}{Ref_gilesmt}{R. Giles, L. McLerran, C. B. Thorn,
{\sl Phys. Rev.} {\bf D17} (1978) 2058.}\fi
\IfSelect \Contents
{-3}{Ref_thornfock}{C. B. Thorn,
{\sl Phys. Rev.} {\bf D20} (1979) 1435.}\fi
\IfSelect \Contents
{-3}{Ref_thornrpa}{C. B. Thorn,
{\sl Phys. Rev.} {\bf D51} (1995) 647.}\fi
\IfSelect \Contents
{-3}{Ref_thornweeparton}{C. B. Thorn,
{\sl Phys. Rev.} {\bf D19} (1979) 639.}\fi
\IfSelect \Contents
{-3}{Ref_puzalowski}{R. Puzalowski,
{\sl Acta Physica Austriaca} {\bf 50} (1978) 45.}\fi
\IfSelect \Contents
{-3}{Ref_feynmanparton}{R. P. Feynman,
Third Topical Conference in High Energy Collisions
of Had\-rons, Stony Brook, N.Y.(1969);
J. D. Bjorken and E. Paschos,
{\sl Phys. Rev.} {\bf 185} (1969) 1975;
J. Kogut and L. Susskind, {\sl Physics Reports} {\bf 8} (1973) 75.}\fi
\IfSelect \Contents
{-3}{Ref_susskindbh}{L. Susskind,
``Some Speculations About Black Hole Entropy in String Theory,''
Rutgers Univ. preprint RU-93-44, hep-th/9309145;
L. Susskind and J. Uglum, ``Black Hole Entropy in Canonical
Quantum Gravity and Superstring Theory,'' Stanford Univ. preprint,
hep-th/9401070;
L. Susskind,
{\sl Phys. Rev.} {\bf D49} (1994) 6606.}\fi
\IfSelect \Contents
{-3}{Ref_bergmantgal}{O. Bergman and C. B. Thorn,
``Super-Galilei Invariant Field Theories in 2 $+$ 1 Dimensions,''}\fi
\IfSelect \Contents
{-3}{Ref_greenssustring}{M. B. Green and J. H. Schwarz,
{\sl Phys. Lett.} {\bf B109} (1982) 444;
{\sl Nucl. Phys.} {\bf B181} (1981) 502;
{\sl Nucl. Phys.} {\bf B198} (1982) 252.}\fi
\IfSelect \Contents
{-3}{Ref_greensb}{M. B. Green, J. H. Schwarz, and L. Brink,
{\sl Nucl. Phys.} {\bf B219} (1983) 437.}\fi
\IfSelect \Contents
{-3}{Ref_greensbook}{M. B. Green, J. H. Schwarz, and E. Witten,
{\it Superstring Theory}, Volumes 1 and 2,
Cambridge University Press (1987).}\fi
\IfSelect \Contents
{-3}{Ref_greenseiberg}{M. B. Green and N. Seiberg,
{\sl Nucl. Phys.} {\bf B299} (1988) 108.}\fi
\IfSelect \Contents
{-3}{Ref_greensiteklink}{J. Greensite and F. R. Klinkhamer,
{\sl Nucl. Phys.} {\bf B281} (1987) 269;
{\sl Nucl. Phys.} {\bf B291} (1987) 557;
{\sl Nucl. Phys.} {\bf B304} (1988) 108.}\fi
\IfSelect \Contents
{-3}{Ref_bergmantbits}{O. Bergman and C. B. Thorn,
``String Bit Models for Superstring,''
preprint UFIFT-HEP-95-8, eprint hep-th/9506125}\fi
\IfSelect \Contents
{-3}{Ref_Beg}{M.A.B. B\'eg and R.C. Furlong,
{\sl Phys. Rev.} {\bf D31} (1985) 1370.}\fi
\IfSelect \Contents
{-3}{Ref_JackiwPi}{R. Jackiw and S.Y. Pi,
{\sl Phys. Rev.} {\bf D42} (1990) 3500;
{\sl Prog. Theor. Phys. Suppl.} {\bf 107} (1992) 1.}\fi
\IfSelect \Contents
{-3}{Ref_pert}{E.L. Feinberg,
{\sl Sov. Phys. Usp. (Eng. Trans.} {\bf 5} (1963) 753;
E. Corinaldesi and F. Rafeli,
{\sl Am. J. Phys.} {\bf 46} (1978) 1185;
K.M. Purcell and W.C. Henneberger,
{\sl Am. J. Phys.} {\bf 46} (1978) 1255.}\fi
\IfSelect \Contents
{-3}{Ref_WittenOlive}{E. Witten and D. Olive,
{\sl Phys. Lett.} {\bf B78} (1978) 97.}\fi
\IfSelect \Contents
{-3}{Ref_Hagen}{C.R. Hagen,
``Perturbation Theory and the Aharonov-Bohm Effect'',
preprint UR-1413, eprint hep-th/9503032.}\fi
\IfSelect \Contents
{-3}{Ref_AB}{Y. Aharonov and D. Bohm,
{\sl Phys. Rev.} {\bf 115} (1959) 485.}\fi
\IfSelect \Contents
{-3}{Ref_Hagengauge}{C.R. Hagen,
{\sl Phys. Rev.} {\bf D31} (1985) 848.}\fi


\def\sgone{{\cal S}_1{\cal G}}
\def\sgtwo{{\cal S}_2{\cal G}}

\def\bx{{\bf x}}
\def\by{{\bf y}}
\def\bz{{\bf z}}
\def\br{{\bf r}}
\def\D{{\cal D}}

\pubnum={$95-12$\cr
hep-th/9507007\cr}
\date{}
\pubtype={}
\titlepage
\title{Super-Galilei Invariant Field Theories in 2+1 Dimensions\foot{Supported
in part by
the Department of Energy under grant DE-FG05-86ER-40272,
and by the Institute for Fundamental Theory.}}
\author{Oren Bergman\foot{E-mail  address: oren@phys.ufl.edu \hfill}
and Charles B. Thorn\foot{E-mail  address: thorn@phys.ufl.edu \hfill}
}
\address{Institute for Fundamental Theory\break
Department of Physics, University of Florida, Gainesville,
FL, 32611, USA }
\abstract
{We extend the Galilei group of space-time transformations by gradation,
construct
interacting field-theoretic representations of this algebra, and show that
non-relativistic Super-Chern-Simons theory is a special case.
We also study the
generalization to matrix valued fields,
which are relevant to the formulation of
superstring theory as a $1/N_c$ expansion of a field theory.
We find that in the
matrix case, the field theory is much more restricted by the supersymmetry.}

\endpage
\pagenumbers
\chapter{Introduction}
Galilean invariance is generally thought of as a low energy approximation
to Poincar\'e invariance, the exact space-time symmetry of relativistic
systems. In this point of view Galilei invariant field theories
describe the dynamics of low energy, or non-relativistic, systems.
There are two main applications for Galilei invariant field theories
in this context. One is to consider them as non-relativistic limits
($c\rightarrow\infty$) of corresponding relativistic field theories,
and to study generic features of field theory in this simpler setting.
This pedagogical approach has been used to exhibit features such as
gauge invariance\[Hagengauge], triviality and renormalization\[Beg],
self-dual solitons\[JackiwPi], and the conformal anomaly\[bergman1] in
non-relativistic field theories, with the
occasional hope of learning something useful about the relativistic theories.

The other application of non-relativistic field theories is
second-quantization of non-relativistic
quantum mechanical systems. This is used in
condensed matter systems, which are non-relativistic by nature.
Frequently a second-quantized approach can shed some light on a seemingly
intractable quantum mechanical problem. A good example is the age old
Aharonov-Bohm (AB) scattering problem\[AB], in which a charged particle
scatters
off an infinitely long and infinitesimally thin solenoid carrying a magnetic
flux. The exact solution was discovered by Aharonov and Bohm in 1959.
Ignoring this fact for a moment, one is naturally led to using
perturbation theory via the Born expansion to get an approximate
solution. Curiously, all attempts
at reproducing even the lowest order term in a Taylor expansion of
the exact solution have failed\[pert], until recently\[bergman2].
In the second-quantized approach of \Refer{bergman2}, one is led
to a natural resolution of the perturbative puzzle by simply including
terms in the action required for consistency of the field theory.

There is an alternative point of view,
in which Galilean invariance is relevant for relativistic
systems. This happens when relativistic systems are quantized in
light-cone variables. Light-cone coordinates are defined by singling
out one of the spatial directions, say $x^{D-1}$, and letting
$$x^\pm = {1\over\sqrt 2}\big(x^0 \pm x^{D-1}\big)\; .$$
In light-cone quantization the role of time is played by $x^+$,
so its conjugate momentum $p^-$ is the light-cone Hamiltonian.
$x^-$ is the longitudinal coordinate, and $x^i$, with
$i=1,\ldots,D-2$, are the transverse coordinates. In these coordinates
a transverse Galilei group in $D-2$ space and one time dimensions
emerges as a manifest
subgroup of the $D$ dimensional Poincar\'e group.
Transverse spatial translations are generated by $p^i$, time translation is
generated by
$p^-$, transverse spatial rotations are generated by the transverse
components of the Lorentz tensor $M^{ij}$, and transverse Galilei boosts
are generated by the mixed components $M^{+i}$. The remaining components
of the Lorentz generator
$M^{-i},M^{+-}$ are not part of the Galilei sub-algebra. From the point
of view of the Galilei subgroup, the longitudinal momentum $p^+$
plays the role of Newtonian
mass, even though it is a generator of the Poincar\'e
group.

One can imagine systems in which the Poincar\'e symmetry breaks down
to its Galilei subgroup in light-cone variables. These systems are
by no means non-relativistic in the usual sense of $c\rightarrow\infty$,
but they are still strictly Galilei invariant. One such system is
relativistic string quantized in light-cone gauge. Except at
$D=26$, where the full Poincar\'e invariance is realized, the dynamics
of light-cone string are only Galilei invariant. In light-cone
gauge $p^+$ is essentially the length of a piece of string. Replacing it
with a discrete variable allows for a description of string
as a composite of string bits, obeying
Galilei invariant dynamics in
$(D-2)+1$ dimensions\[gilest,klebanovs,thornmosc].
The mass of each bit is $p^+/({\rm \# bits})$, so the total mass is
$p^+$, as implied by the Galilei algebra. The dynamics of the bits must
be such as to give a strong nearest neighbor attraction and weaker
non-nearest neighbor interactions, in order for long closed polymers
to form.
The missing dimension
of string, the coordinate $x^-$, reappears in the limit
where its conjugate $p^+$
becomes continuous, and the discrete polymer becomes a continuous string.
The dynamics of the bits are described by a Galilei invariant field
theory. But a nearest neighbor interaction pattern in continuous space
clearly cannot be achieved by single component fields. It requires
the introduction of a matrix field theory and the use of 't Hooft's
$1/N_c$ expansion\[thornmosc]. The nearest neighbor
interaction will then appear at
zeroth order in this expansion. Non-nearest neighbor interactions, which lead
to the breaking of a polymer into several polymers, will appear
at higher orders in the expansion.

The possibility of extending the Galilei group of transformations
by a gradation to a Super-Galilei algebra in $3+1$ dimensions
was first suggested
in \Refer{puzalowski}. The author showed that there are two
possible superalgebras $\sgone$ and $\sgtwo$, where $\sgone\subset\sgtwo$.
The smaller superalgebra includes a single two-component spinor
supercharge $Q$, and the larger superalgebra includes in addition
a second two-component spinor supercharge $R$. He then proceeded to
construct field theoretic representations of the $\sgone$ algebra.
\Refer{MIT} explored a particular $\sgtwo$ invariant field theory
in $2+1$ dimensions, namely non-relativistic Super-Chern-Simons theory.
It was actually derived as a non-relativistic limit of relativistic
Super-Chern-Simons theory\[supercs]. The authors showed that the
existence of non-relativistic self-dual solitons was guaranteed by
supersymmetry. Later\[bergman2] it was also suggested that the conformal
anomaly of the non-relativistic Chern-Simons theory vanishes in the
supersymmetric case, bringing together the concepts of supersymmetry,
self-duality and conformal invariance.

As with the Galilei algebra, the Super-Galilei algebra appears
as a subalgebra of the Super-Poincar\'e algebra in light-cone
coordinates. The possibility of constructing Super-Galilei invariant
field theories may then lead to a reformulation of superstring theory
as a supersymmetric bit theory in one less dimension. We leave this issue
for another paper\[bergmantbits]. In this paper we are interested in
constructing $\sgtwo$ invariant field theories in $2+1$ dimensions.
It is this larger superalgebra that emerges as a subalgera of the
Super-Poincar\'e algebra, whereas the non-relativistic limit of the
Super-Poincar\'e algebra is just $\sgone$. We consider only $2+1$ dimensions
since that corresponds to four dimensional Super-Poincar\'e invariance.
Critical superstring lives in ten dimensions, and thus the bit
formulation should be in $8+1$ dimensions. $2+1$ dimensional
Super-Galilei invariant field theories serve first of all as toy models
for the $8+1$ dimensional model we eventually want to construct. In
addition, they may also be the basis of a physical four dimensional
superstring theory, with compactified dimensions built out of internal
degrees of freedom in the bit theory.

In section 2 we present the Super-Galilei algebras $\sgone$ and
$\sgtwo$ in $2+1$ dimensions. In section 3 we construct field
theoretic representations for $\sgtwo$ by second-quantizing all
the charges and deriving the Hamiltonian from the superalgebra.
We also discuss the Super-Fock space and the super-wavefunctions.
In section 4 we show that non-relativistic Super-Chern-Simons theory
is just a special case of the general Galilei invariant field theory
constructed in section 3. In section 5 we construct a Super-Galilei
invariant matrix field theory, and discuss singlet Fock states
and the $1/N_c$ expansion. This section is a prelude to developing
a bit model for superstring in the Green-Schwarz formulation.
In the last section
we present a brief discussion of our results.

\chapter{The Super-Galilei Algebra in 2+1 Dimensions}
The generators of the Galilei group in 2+1 dimensions include
a 2 dimensional momentum vector ${\bf P}$, a 2 dimensional boost
vector ${\bf K}$, a planar angular momentum scalar $J$, and a
Hamiltonian $H$. In addition there is also a number operator $M$,
counting the total number of particles in the system. The
only non-vanishing commutators in the algebra are given by
$$\eqalign{
 [P_i, K_j] &= i\delta_{ij}mM \cr
 [H,K_i] &= iP_i \cr}\qquad\eqalign{
 [P_i,J] &= -i\epsilon_{ij}P_j \cr
 [K_i,J] &= -i\epsilon_{ij}K_j \; .\cr}
\(Galilei)$$
One can extend this algebra by adding a complex odd (fermionic) charge $Q$,
satisfying the following commutator:
$$
[Q,J] = {1\over 2}Q \; ,
\(supercom)$$
and its hermitian conjugate counterpart. All other commutators vanish.
The graded algebra will close if in addition the following anti-commutators
(and their hermitian conjugates) are satisfied:
$$\eqalign{
 \{Q,Q^\dagger\} &= \vphantom{1\over 2}mM \cr
 \{Q,Q\} &= 0 \; .\cr}
\(Qalgebra)$$
The graded algebra given by \(Galilei)-\(Qalgebra) defines
the superalgebra $\sgone$.
The extended Super-Galilei algebra $\sgtwo$ requires an additional
supercharge $R$ satisfying the following commutators:
$$\eqalign{
  [R,J] &= -R/2\cr
  [R,K^-] &= -iQ \; , \cr}
\(Rcoms)$$
and their hermitian conjugates, with all other commutators vanishing.
The $\pm$ components of any real two dimensional vector ${\bf V}$ are
defined by $V^{\pm}\equiv V^1 \pm iV^2$.
To close the $\sgtwo$ algebra we need the following anti-commutators:
$$\eqalign{
 \{Q,R\} &= \{R,R\} = 0 \cr
 \{Q,R^\dagger\} &= -P^-/2 \cr
  \{R,R^\dagger\} &= H/2 \; ,\cr}
\(Ralgebra)$$
and their hermitian conjugates.
The algebra given by \(Galilei)-\(Ralgebra) then defines the
super-algebra $\sgtwo$. We turn next to a field theoretic representation
of this super-algebra.

\chapter{Field Theoretic Representation of $\sgtwo$}
The simplest $N=1$ Galilei supermultiplet in 2+1 dimensions consists
of a complex scalar
field $\phi(\bx)$ and a one-component complex Grassmann field $\psi(\bx)$
corresponding to a spin helicity of $-1/2$ in the plane.
The fields satisfy the canonical commutation relations:
$$
 [\phi(\bx),\phi^\dagger(\by)] = \{\psi(\bx),\psi^\dagger(\by)\} =
\delta(\bx-\by)\; .
$$
The superalgebra $\sgtwo$ can then be realized with free fields
as follows:
$$\eqalign{
 M &= \int d\bx \big[\phi^\dagger(\bx)\phi(\bx) +
\psi^\dagger(\bx)\psi(\bx)\big] \cr
 P^i &= -i\int d\bx \big[\phi^\dagger(\bx)\partial^i \phi(\bx) +
                        \psi^\dagger(\bx)\partial^i \psi(\bx)\big] \cr
 K^i &= -\int d\bx \Big[\phi^\dagger(\bx)
\big(it\partial^i + mx^i\big)\phi(\bx)
+         \psi^\dagger(\bx)\big(it\partial^i + mx^i\big)\psi(\bx)\Big] \cr
 J &= -i\int d\bx \Big[\phi^\dagger(\bx)\big(\bx\times
          \nabla\big)\phi(\bx) +
   \psi^\dagger(\bx)\big(\bx\times
          \nabla - {i\over 2}\big)\psi(\bx)\Big] \cr
 {\cal Q} &= -i\sqrt{m}\int d\bx \psi^\dagger(\bx)\phi(\bx) \cr
 {\cal Q}^\dagger &= i\sqrt{m}\int d\bx \phi^\dagger(\bx)\psi(\bx) \cr
 {\cal R}^{(0)} &= {1\over 2\sqrt{m}}
\int d\bx \psi^\dagger(\bx)\partial^+\phi(\bx) \cr
 {\cal R}^{(0)\dagger} &= -{1\over 2\sqrt{m}}\int d\bx
\phi^\dagger(\bx)\partial^-\psi(\bx)\; . \cr}
\(freecharges)$$
We use script letters for the second-quantized supercharges to avoid
later confusion with their first-quantized counterparts.
The Hamiltonian for this field theory is then clearly
$$
 H^{(0)} = 2\{{\cal R}^{(0)},{\cal R}^{(0)\dagger}\} =
{1\over 2m}\int d\bx \big[|\nabla\phi(\bx)|^2 +
|\nabla\psi(\bx)|^2\big] \; .
\(freehamiltonian)$$
To construct an interacting field theory one usually adds higher order terms
to the Hamiltonian (or action). To check that the resulting theory is
supersymmetric is somewhat cumbersome, it is more convenient to
add higher order terms to the
supercharge ${\cal R}^{(0)}$ instead. For the resulting theory
to be supersymmetric certain conditions on the higher order terms
must hold.
Let ${\cal R}^\prime$ denote the additional terms, then the total supercharge
is
$${\cal R} = {\cal R}^{(0)} + {\cal R}^\prime \; , \(totalR)$$
and the total Hamiltonian is given by
$$H = 2\{{\cal R},{\cal R}^\dagger\} \; . \(totalH)$$
This supercharge must satisfy the $\sgtwo$ algebra,
and consequently ${\cal R}^\prime$ must satisfy the following relations
$$\eqalign{
  [{\cal R}^{\prime},M] &= 0 \cr
  [{\cal R}^\prime,K^{\pm}] &= 0 \cr
  [{\cal R}^\prime,J] &= -{1\over 2} {\cal R}^\prime\cr}
 \qquad\eqalign{
  \{{\cal R}^\prime,{\cal Q}\} &= 0\cr
  \{{\cal R}^\prime,{\cal Q}^\dagger\} &= 0 \cr
  2\{{\cal R}^\prime,{\cal R}^{(0)}\} +
     \{{\cal R}^\prime,{\cal R}^\prime\} &= 0\; . \cr}
\(restrictions)$$
The conjugate supercharge ${\cal R}^{\prime\dagger}$ satisfies
similar relations, except the
spin is reversed.
Invariance under the global $U(1)$ symmetry given by the first
commutator implies that ${\cal R}^\prime$ has an equal number of creation
and annihilation operators. Invariance under Galilei boosts given
by the second commutator further
restricts this to be so at each point.
For simplicity we limit modifications to quartics in the fields.
The anti-commutation relations with
${\cal Q}$ and ${\cal Q}^\dagger$ then restrict the form of
${\cal R}^\prime$ to:
$$\eqalign{
 {\cal R}^\prime & \propto \int
d\bx\,d\by\,V^+(\by-\bx)\psi^\dagger(\bx)\rho(\by)\phi(\bx)\cr
 {\cal R}^{\prime\dagger} & \propto \int
d\bx\,d\by\,V^-(\by-\bx)\phi^\dagger(\bx)\rho(\by)\psi(\bx)
  \; , \cr}
\(general)$$
where $\rho = \phi^\dagger \phi + \psi^\dagger\psi$. Finally,
the spin condition restricts the functions $V^{\pm}$ to be
of the following form,
$$\eqalign{
 V^+(\bx) &= (\partial^1 + i\partial^2)f(|\bx|)\cr
 V^-(\bx) &= (\partial^1 - i\partial^2)f^*(|\bx|)\; .\cr}
\(constraint)$$
It is now straightforward to show that
the above supercharges satisfy the last of the conditions
in \(restrictions).
The total supercharges can be written concisely as
$$\eqalign{
  {\cal R} &= {1\over 2{\sqrt m}}\int d\bx \psi^\dagger(\bx)\D^+\phi(\bx) \cr
  {\cal R}^\dagger &= -{1\over 2{\sqrt m}}\int d\bx \phi^\dagger(\bx)\D^
  -\psi(\bx)\; , \cr}
\(intcharges)$$
where
$$
 \D^{\pm} = \partial^{\pm} - i\int d\by V^{\pm}(\by - \bx)\rho(\by)\; .
\(covariantd)$$
The transformation of the component fields under the
$\sgtwo$ algebra can be read off from \(freecharges) and
\(intcharges), by taking
commutators of the fields with the charges.

The Hamiltonian obtained by anticommuting the supercharges in
\(intcharges) is
given by
$$\eqalign{
 H &={1\over 2m}\int d\bx\, \big[|\nabla\phi(\bx)|^2 +
|\nabla\psi(\bx)|^2\big]\cr
   &+{i\over 2m}\int d\bx\,
   d\by\,\Big[V^+(\by-\bx)\big(-\partial^-\phi^\dagger(\bx)\rho(\by)
     \phi(\bx) + \psi^\dagger(\bx)\rho(\by)\partial^-\psi(\bx)\big) - {\rm
h.c.}\Big]\cr
  & -{i\over 2m}\int
d\bx\,d\by\,\Big[\partial_y^-V^+(\by-\bx)\psi^\dagger(\bx)
     \phi^\dagger(\by)\phi(\bx)\psi(\by) - {\rm
h.c.}\Big] \cr
  &+{1\over 2m}\int d\bx\,d\by\,d\bz\, V^+(\by-\bx)V^-(\bz-\bx)\big[
     \phi^\dagger(\bx)\rho(\bz)\rho(\by)\phi(\bx) +
\psi^\dagger(\bx)\rho(\by)\rho(\bz)\psi(\bx)
     \big]\; .\cr}
\(hamiltonian)$$
The above Hamiltonian defines a Super-Galilei ($\sgtwo$)
invariant quantum field theory.
The Fock space of this field theory consists of bosonic and fermionic states
created by $\phi^\dagger$ and $\psi^\dagger$, respectively. The two creation
operators can be collected into a single superfield,
$$\Phi^\dagger(\bx,\theta) = \phi^\dagger(\bx) + \psi^\dagger(\bx)\theta\; ,$$
where $\theta$ is an anti-commuting c-number. This field creates a single
``superparticle''. Multi-superparticle states are created by acting on the
vacuum with several superfields,
$$\ket{\Psi}=\int\prod_{k=1}^M\big(d^{2}x_kd\theta_k\big)
\Phi^\dagger({\bf x}_1\theta_1)\cdots\Phi^\dagger({\bf x}_M\theta_M)\ket{0}
\Psi({\bf x}_1\theta_1,\cdots,{\bf x}_M\theta_M)\; .
\(Mstate)
$$
The super-wavefunction $\Psi$ is composed of component wave functions,
each of which describes a well defined number of bosons and
a well defined number of fermions.

By acting on the state $\ket{\Psi}$ with the generators of $\sgtwo$,
one can derive the
first-quantized representations of these generators which act
on the super-wavefunction. In particular, the supercharges are
given by:
$$\eqalign{
 Q &= -i\sqrt{m}\sum_{k=1}^M{\partial\over\partial\theta_k}
 \qquad , \qquad Q^\dagger = i\sqrt{m}\sum_{k=1}^M\theta_k \cr
 R & = {1\over 2\sqrt m}\sum_{k=1}^M\Big[\partial^+_k - i\sum_{l\neq k}
         V^+(\bx_l-\bx_k)\Big]{\partial\over\partial\theta_k} \cr
 R^\dagger & = -{1\over 2\sqrt m}\sum_{k=1}^M\Big[\partial^-_k - i\sum_{l\neq
k}
         V^-(\bx_l-\bx_k)\Big]\theta_k \; .\cr}
\(firstquantized)$$
By anti-commuting $R$ and $R^\dagger$, or equivalently
using the quantum field equations of motion,
$$i\partial_t\Phi^\dagger = [\Phi^\dagger,H]\; ,$$
in \(Mstate), we arrive at the first-quantized form of the Hamiltonian and
the Schr\"odinger equation for the super-wavefunction,
$$\eqalign{
 i\partial_t\Psi = \bigg\{& -{1\over 2m}\sum_k\nabla_k^2
   +{i\over m}\sum_{k,l\neq k}{\bf V}(\bx_l-\bx_k)\cdot\nabla_k \cr
 & +{i\over 2m}\sum_{n,k,l\neq k}\Big[\partial_n^+V^-(\bx_l-\bx_k)
             {\partial\over\partial\theta_n}\theta_k +
 \partial_n^-V^+(\bx_l-\bx_k)\theta_n{\partial\over\partial\theta_k}\Big] \cr
 & + {1\over 2m}\sum_{k,l\neq k,n\neq k}V^-(\bx_l-\bx_k)
     V^+(\bx_n-\bx_k)\bigg\}\Psi\; , \cr}
\(Schrodinger)$$
where ${\bf V}=({\rm Re}V^\pm,\pm{\rm Im}V^\pm)$.

\chapter{Non-relativistic Super-Chern-Simons Theory}
We begin by
using \(covariantd)\ and some integration by parts to rewrite
the Hamiltonian in the following suggestive manner,
$$\eqalign{
 H =& {1\over 2m}\int d\bx\,\big[|\D^+\phi(\bx)|^2 +
|\D^+\psi(\bx)|^2\big]\cr
 &+{1\over m}\int d\bx\,d\by\,\nabla_y\times{\bf
V}(\by-\bx)\big[\psi^\dagger(\bx)
   \phi^\dagger(\by)\psi(\by)\phi(\bx) -
\psi^\dagger(\bx)\rho(\by)\psi(\bx)\big]\; .\cr}
\(conciseham)$$
Note that for the special choice
\foot{For vectors in the plane the cross product is defined by
     ${\bf V}\times{\bf U}=\epsilon^{ij}V^iU^j$, and the curl
     is defined by $\nabla\times{\bf V}=\epsilon^{ij}\partial^iV^j$.
     The curl of a scalar is defined by $(\nabla\times S)^i=
     \epsilon^{ij}\partial^jS$. This quasi three dimensional
     vector notation makes sense because, in
     dimensional reduction from 3 dimensions, the ``3'' component
     is an $SO(2)$ scalar. Thus $\nabla\times{\bf V}$ has only a ``3''
     component and is a scalar. Similarly $\nabla\times S$ for an $SO(2)$
     scalar function $S$ is to be thought of as the curl of a 3-vector
     with only 3rd component non-vanishing and equal to $S(x^1,x^2)$.}
$${\bf V}(\bx) = \alpha\nabla\times\ln|\bx| \; ,$$
with $\alpha$ an arbitrary real constant,
one gets $\nabla\times{\bf V}(\bx)
=-2\pi\alpha\delta(\bx)$, and only the top part
of \(conciseham)\ remains. Such a theory is characterized by static classical
configurations (solitons) obeying a first order (self-dual) differential
equation,
$$\D^+\phi = \D^+\psi = 0 \; .\(selfdual)$$
We will have more to say about this later.

As the notation suggests, we can interpret $\D$ as a covariant derivative,
$\D = \nabla - ie{\bf A}$, where
${\bf A}$ is a ``background abelian gauge field'' given by
$$ {\bf A} = {1\over e} \int d\by\,{\bf V}(\by-\bx)\rho(\by)\; .
\(gauge)$$
The following identities
$$\eqalign{
 \int d\bx\,|\D^+\phi|^2 &= \int d\bx\,\Big[|\D\phi|^2 + \int d\by\,
     \nabla_y\times{\bf V}(\by-\bx)
     \phi^\dagger(\bx)\rho(\by)\phi(\bx)\Big]\cr
 \int d\bx\,|\D^+\psi|^2 &= \int d\bx\,\Big[|\D\psi|^2 + \int d\by\,
     \nabla_y\times{\bf V}(\by-\bx)
     \psi^\dagger(\bx)\rho(\by)\psi(\bx)\Big]\; ,
\cr}
\(identity)$$
then allow us to express
the Hamiltonian as a minimal coupling of the matter to the
``gauge field'' plus additional matter coupling terms,
$$\eqalign{
 H = {1\over 2m}\int d\bx\,\big[|{\D}\phi|^2 &+ |{\D}\psi|^2\big]
            +{1\over 2m}\int d\bx\,d\by\,\Big[\nabla_y\times{\bf
V}(\by-\bx)\Big] \cr
    & \times\Big[:\big(|\phi(\bx)|^2 - |\psi(\bx)|^2\big)\rho(\by):
        + 2\psi^\dagger(\bx)\phi^\dagger(\by)\psi(\by)\phi(\bx)\Big]\; .\cr}
\(minimal)$$

For a specific choice of the vector function ${\bf V}(\bx)$ the Hamiltonian
\(minimal)\ can be derived by solving
the Gauss' law constraint of a particular
Super-Galilei invariant {\em gauge theory}, namely non-relativistic
Super-Chern-Simons theory.
This is the only known example of a Super-Galilei invariant
gauge theory\[MIT].
Let us review the construction of this theory.
Chern-Simons theory coupled to nonrelativistic
bosons and fermions is described by
the following action:
$$\eqalign{
S_{\rm CS} = \int & d^3x\,\Big[ {\kappa\over 2}\partial_t{\bf A}\times{\bf A} -
\kappa A^0 B
  + \phi^\dagger\Big(i\D_t + {\D^2\over 2m}\Big)\phi
   + \psi^\dagger\Big(i\D_t + {\D^2\over 2m}\Big)\psi\cr
  &- {e\over 2m}B|\psi|^2 + \lambda_1 |\phi|^4 + \lambda_2 |\phi|^2
|\psi|^2\Big]\; ,\cr}
\(csaction)$$
where the time component of the covariant derivative is given by
$\D_t = \partial_t + ieA^0$. The Pauli interaction term has been
explicitly
included, as well as two additional matter coupling terms with coupling
constants
$\lambda_1, \lambda_2$. The theory possesses
the $\sgtwo$ Super-Galilei symmetry
for the following values of the coupling constants:
$$\lambda_1 = - {e^2\over 2m\kappa} \qquad, \qquad\lambda_2 = 3\lambda_1
\;.\(couplings)$$
Note that the last three terms in \(csaction)\ differ by minus signs from
the same terms in \Refer{MIT},
since our convention
for the helicity of the fermion is opposite to theirs.
The normal ordered Hamiltonian derived from \(csaction)\ is given by
$$\eqalign{
H_{\rm CS} = {1\over 2m} \int d\bx\,&\Big[|\D\phi|^2 + |\D\psi|^2\cr
       & + e:B|\psi|^2: - 2m\lambda_1:|\phi|^4: - 2m\lambda_2 |\phi|^2
|\psi|^2\Big]\; .\cr}
\(cshamiltonian)$$
This Hamiltonian is accompanied by the Gauss' Law constraint,
derived by varying
\(csaction)\ with respect to $A_0$,
$$B = -{e\over\kappa}\big(\phi^\dagger\phi + \psi^\dagger\psi\big)\; .
\(gauss)$$
The solution of this constraint in Coulomb gauge is given by
$${\bf A}(\bx) = -{e\over\kappa}
      \int d\by\,\big[\nabla_y\times\ln|\by-\bx|\big]\rho(\by)\; .\(csfield)$$
Consequently the Super-Chern-Simons Hamiltonian $H_{\rm CS}$
agrees with the Hamiltonian
in Eq.\(minimal) for
${\bf V}(\bx) = -(e^2/\kappa)\nabla\times\ln|\bx|$.

Interestingly, this is the same vector function for which
the Hamiltonian had a self-dual form. In fact the non-relativistic
Super-Chern-Simons
theory does indeed possess self-dual solitons\[MIT],
which are generalizations of
the non-relativistic Chern-Simons solitons discovered by
Jackiw and Pi\[JackiwPi]. In
the
purely bosonic theory, self-duality was imposed by hand, by adding a contact
interaction term of appropriate strength. In the supersymmetric case
self-duality
is automatic. The connection between supersymmetry and self-duality in
relativistic
field theories has been known for quite a while\[WittenOlive]. The above
analysis
indicates that there is a connection between supersymmetry and self-duality in
some Galilei invariant (non-relativistic) theories as well.

In addition to $\sgtwo$ invariance, the non-relativistic Super-Chern-Simons
theory
\(csaction)\ also possesses an $SO(2,1)$ conformal invariance\[MIT]. Such a
Galilean
conformal symmetry is usually broken by quantum mechanical
anomalies\[bergman1,bergman2],
but it turns out that supersymmetry guarantees
that it is anomaly free\[bergman2].
Thus supersymmetry, self-duality and conformal invariance co-exist
at the particular point in parameter space given by \(couplings).
If the parameters $\lambda_1, \lambda_2$ are changed, not only
would it spoil supersymmetry, but also self-duality and conformal
invariance.

\chapter{Matrix Valued Fields}
As one of the motivations for studying Super-Galilei invariant
field theories,
we mentioned that superstrings and
their interactions may be a consequence of a
$1/N_c$ expansion of a Super-Galilei invariant
unitary matrix field theory. In a separate
paper\[bergmantbits]
we construct a field theory that gives the free superstring. The
point-like
objects (bits) of the field theory carry two color indices, and are created by
the $N_c\times N_c$ matrix valued fields $\phi^\dagger(\bx)_\alpha^\beta$ and
$\psi^\dagger(\bx)_\alpha^\beta$.
The canonical commutators of the matrix fields are given by
$$[\phi(\bx)_\alpha^\beta,\phi^\dagger(\by)_\gamma^\delta] =
  \{\psi(\bx)_\alpha^\beta,\psi^\dagger(\by)_\gamma^\delta\} = \delta(\bx-\by)
  \delta_\alpha^\delta\, \delta_\gamma^\beta\; .\(matrixcom)$$
In addition to Super-Galilean invariance
(either $\sgone$ or $\sgtwo$) the field theory is required to have a global
$U(N_c)$ symmetry. The fields are matrices
transforming in the adjoint representation of
$U(N_c)$, and
the terms in the action, or Hamiltonian,
involve traces of
products of matrices. The generalization of the free $\sgtwo$ charges
\(freecharges)
to matrix fields is straightforward: simply elevate the fields to matrix
fields, understanding all products as matrix products,
and take the trace. This is true for any operator which is quadratic in the
fields.
The free Hamiltonian is then just the trace of \(freehamiltonian).

For products of more than two fields the matrix ordering (color routing)
is important, since
different orderings (routings) can lead to different traces.
By the cyclicity of the
trace, there are $(n-1)!$ ways to order $n$ matrix fields inside a trace. In
particular there are thirty six possibilities for the interaction term in the
supercharge ${\cal R}$.
However the $\sgtwo$ algebra is only satisfied for some of them, and only
for a special choice of the function $V^+(\bx)$. Consider for example
the following supercharge:
$${\cal R}_1^\prime = {-i\over 2N_c\sqrt m}\int d\bx\,d\by
                     \,V^+(\by-\bx):\Tr\big[
          \psi^\dagger(\bx)\phi(\bx)\rho(\by)\big]:\; ,
\(R1)$$
where
$\rho_\alpha^\beta = [\phi^\dagger\phi + \psi^\dagger\psi]_\alpha^\beta$,
and $V^+(\bx)=\partial^+f(|\bx|)$ as before.
It satisfies all but the last equation in \(restrictions),
$$\eqalign{
 2\{{\cal R}_1^\prime,&{\cal R}^{(0)}\} + \{{\cal R}_1^\prime,{\cal
R}_1^\prime\}
  = {-1\over 2mN_c^2} \int d\bx\,d\by\,d\bz
    :\Tr\Big[\rho(\bx)\psi^\dagger(\by)\phi(\by)\psi^\dagger(\bz)\phi(\bz)
        \Big]: \cr
 &\times\Big[V^+(\bx-\bz)V^+(\bz-\by)+V^+(\bx-\by)V^+(\by-\bz)+
     V^+(\bz-\bx)V^+(\bx-\by)\Big] \; , \cr}
\(anticom)$$
which vanishes
only when the
function $V^+(\bx)$ satisfies
$$V^+(\bx-\bz)V^+(\bz-\by)
+V^+(\bx-\by)V^+(\by-\bz)+V^+(\bz-\bx)V^+(\bx-\by) = 0 \; .
$$
When combined with the constraint \(constraint),
$V^+(\bx)=\partial^+f(|\bx|)$,
the solution to the above condition is
$$V^+(\bx) = \alpha\,\partial^+\ln|\bx| \; ,\(solution)$$
where $\alpha$ is an arbitrary complex number, $\alpha=\alpha_1 - i\alpha_2$.
In the field theory of section 3
there was no restriction on $V^+(\bx)$ other than
\(constraint). The requirement of $\sgtwo$ supersymmetry in the
matrix field theory
restricts this function much more.
The Hamiltonian is again found
by anti-commuting the total supercharge ${\cal R}$ with
its hermitian conjugate. We refrain from presenting its explicit form
due to its length.

The Fock space of this theory consists of states transforming in
various representations of $U(N_c)$. As we are primarily interested
in applying matrix field theories to a reformulation of
superstring theory, let us restrict our discussion to the singlet
states given by products of matrix traces of products of
creation operators acting
on the vacuum. Single trace states are defined as
$$\ket{\Psi}=\int\prod_{k=1}^M\big(d^{2}x_kd\theta_k\big)
\Tr[\Phi^\dagger({\bf x}_1\theta_1)\cdots\Phi^\dagger({\bf
x}_M\theta_M)]\ket{0}
\Psi({\bf x}_1\theta_1,\cdots,{\bf x}_M\theta_M)\; ,
\(stringfock)
$$
where $\Psi$ is the wavefunction describing a closed chain of $M$
bits in a first-quantized formalism. Acting on this state with
the supercharge ${\cal R}$ one finds that the trace structure
is altered, and thus it cannot be an energy eigenstate.
Singlet operators like ${\cal Q}$
and ${\cal R}$ relate
singlet states to other singlet states, so a single chain can in
general break into several chains. One-body operators always preserve
the number of traces, so a state of the form \(stringfock) is changed
to a state of the same form by
${\cal Q}$ and ${\cal R}^{(0)}$. Two-body operators such as
${\cal R}^\prime$ can change
the number of traces by one. However, if the matrix ordering
in a two-body operator is such that the creation operators are
{\bf consecutive} there will be terms
in which the number
of traces doesn't change,
and they will get multiplied by a factor of $N_c$. To see
how this happens consider for simplicity a single component matrix
creation operator
$a^\dagger(x)_\alpha^\beta$, and let $\Omega_2$ be a single trace
2-body operator with consecutive creation operators,
$$\Omega_2={1\over N_c}\int dxdyV(y-x)\Tr[a^\dagger(x)a^\dagger(y)a(y)a(x)]\; .
\(twobody)$$
Applying this operator to the
singlet Fock state
$\ket{M}=\Tr[a^\dagger(x_1)\cdots a^\dagger(x_M)]\ket{0}$,
gives after one contraction
$$\displaylines{
\Omega_2\ket{\psi}={1\over N_c}\int dy\sum_k V(y-x_k)\cdot\hfill\cr
\hfill\cdot\Tr\big[ a^\dagger(x_k)a^\dagger(y)a(y)a^\dagger(x_{k+1})\cdots
a^\dagger(x_M)
a^\dagger(x_1)\cdots a^\dagger(x_{k-1})\big]\ket{0}\; .
\cr}$$
To continue the evaluation
we note that it matters crucially which creation operator the last
remaining $a(y)$ contracts against.
The contraction with $a^\dagger(x_{k+1})$ produces
a factor of $\sum_\alpha\delta_\alpha^\alpha=N_c$.
All other contractions fail to provide this factor.
Thus {\bf in the limit $N_c\rightarrow\infty$}
$$\Omega_2\Tr[a^\dagger(x_1)\cdots a^\dagger(x_M)]\ket{0}
\rightarrow\sum_{k=1}^M V(x_{k+1}-x_k)
\Tr[a^\dagger(x_1)\cdots a^\dagger(x_M)]\ket{0}\; .
\(firstquan)$$
Note that only a nearest neighbor interaction survives once we take
the limit $N_c\rightarrow\infty$. This is precisely what we
require of a non-interacting polymer chain of bits.
The other contractions change the trace structure of
the state, giving $1/N_c$ times a state with two traces. Thus $1/N_c$
corrections allow a closed polymer chain to rearrange its
bonds and transform to two closed polymer chains.

The matrix ordering of the supercharge ${\cal R}_1^\prime$ in \(R1)
is such that there are no consecutive annihilation operators, and the
same will be true of the Hamiltonian. A nearest neighbor interaction
pattern will thus not be established in the limit
$N_c\rightarrow\infty$. We therefore seek other possibilities for
${\cal R}^\prime$ to remedy this situation. Consider the following
ordering:
$$
{\cal R}_2^\prime = {-i\over 2N_c\sqrt m}\int d\bx\,d\by
                     \,V^+(\by-\bx):\Tr\big[
          \psi^\dagger(\bx)\rho(\by)\phi(\bx)\big]: \; .
\(r2)
$$
The above contains consecutive annihilation operators, but
fails to anti-commute with
${\cal Q}^\dagger$ for {\bf any} non-trivial function $V^+(\bx)$,
$$\{{\cal R}_2^\prime,{\cal Q}^\dagger\} = {1\over 2N_c}
\int d\bx\,d\by\,V^+(\by-\bx)
:\Tr\big[\big(\phi(\bx)\phi^\dagger(\bx)
-\psi(\bx)\psi^\dagger(\bx)\big)\rho(\by
)\big]:
\; \neq 0 \; .
\(fail)$$
What is needed is a more complicated ordering than ${\cal R}_1^\prime$
or ${\cal R}_2^\prime$, which contains consecutive annihilation
operators {\bf and} satisfies the entire $\sgtwo$ algebra.
The following
combination\foot{The commutators and anti-commutator above refer
only to matrix ordering (color routing), whereas normal ordering refers to the
operator elements of the matrices.}
$${\cal R}^\prime = {-i\over 2N_c\sqrt m}\int d\bx\, d\by\,
       V^+(\by-\bx):\Tr\Big[\Big(
       [\phi^\dagger(\by),\phi(\by)] + \{\psi^\dagger(\by),\psi(\by)\}\Big)
       [\psi^\dagger(\bx),\phi(\bx)]\Big]: \; ,\(intmatcharge)$$
contains consecutive annihilation operators and satisfies all but
the last equation in \(restrictions),
$$\eqalign{
 2\{{\cal R}^\prime,&{\cal R}^{(0)}\} + \{{\cal R}^\prime,{\cal R}^\prime\}
  = {-1\over 4mN_c^2} \int d\bx\,d\by\,d\bz \cr
 &\times :\Tr\Big[\Big([\phi^\dagger(\bx),\phi(\bx)]
              + \{\psi^\dagger(\bx),\psi(\bx)\}\Big)
               [\psi^\dagger(\by),\phi(\by)]
               [\psi^\dagger(\bz),\phi(\bz)]\Big]: \cr
 &\hskip-3pt\times\Big[V^+(\bx-\bz)V^+(\bz-\by)+V^+(\bx-\by)V^+(\by-\bz)+
     V^+(\bz-\bx)V^+(\bx-\by)\Big] \; , \cr}
\(anticom2)$$
which again vanishes only for $V^+(\bx)=\alpha\partial^+\ln|\bx|$.

The first-quantized representations of the supercharges are
obtained by acting on the state $\ket{\Psi}$ and taking the limit
$N_c\rightarrow\infty$,
$$\eqalign{
 Q &= -i\sqrt{m}\sum_{k=1}^M{\partial\over\partial\theta_k}
 \qquad , \qquad Q^\dagger  = i\sqrt{m}\sum_{k=1}^M\theta_k \cr
 R & = {1\over 2\sqrt m}\sum_{k=1}^M\Big[\partial^+_k
        + i\big(V^+(\bx_{k-1}-\bx_k)-V^+(\bx_k-\bx_{k+1})\big)
           \Big]{\partial\over\partial\theta_k} \cr
 R^\dagger & = -{1\over 2\sqrt m}\sum_{k=1}^M\Big[\partial^-_k
           + i\big(V^-(\bx_{k-1}-\bx_k)-V^-(\bx_k-\bx_{k+1})\big)
           \Big]\theta_k \; , \cr}
\(firstquanmatrix)$$
where $V^-(\bx)=\alpha^*\partial^-\ln|\bx|$.
The difference between these and the supercharges in the non-matrix
case \(firstquantized) is that the two body terms in $R$ and $R^\dagger$
include only {\bf nearest neighbor} interactions.
By taking the anti-commutator of $R$ and $R^\dagger$ we arrive at the
first-quantized Hamiltonian:
$$\eqalign{
H =& -{1\over 2m}\sum_{k=1}^M\nabla_k^2
       +{i\over m}\sum_{k=1}^M{\bf V}(\bx_{k}-\bx_{k+1})
       \cdot\big(\nabla_{k}-\nabla_{k+1}\big)    \cr
   & + {i\over 2m}\sum_{k=1}^M\Big\{
      2\partial_k^+V^-(\bx_k-\bx_{k+1})  \cr
   & \qquad-\big[\partial_k^+V^-(\bx_k-\bx_{k+1})-
      \partial_k^-V^+(\bx_k-\bx_{k+1})\big]
       \big(\theta_{k+1}-\theta_k\big)\Big({\partial\over\partial\theta_{k+1}}
      - {\partial\over\partial\theta_k}\Big)\Big\} \cr
    & + {1\over 2m}\sum_{k=1}^M \Big\{ 2|{\bf V}(\bx_k-\bx_{k+1})|^2 \cr
 & \qquad - V^+(\bx_{k-1}-\bx_k)V^-(\bx_k-\bx_{k+1})
          - V^+(\bx_k-\bx_{k+1})V^-(\bx_{k-1}-\bx_k)\Big\}\; .\cr}
\(bigham)
$$
Recall that $V^\pm=V^1\pm iV^2$ and $\alpha=\alpha_1-i\alpha_2$, therefore
$${\bf V}(\bx) = \alpha_1\nabla\ln|\bx| + \alpha_2\nabla\times\ln|\bx|\; .
\(vectorfun)$$

The above Hamiltonian with the vector function \(vectorfun) describes
supersymmetric dynamics of bits which are ordered around a loop.
This does not
yet imply that this loop is in any sense a physical bound chain
(see \Figure{loop}).
That question must be answered by studying the bound states of the
system, if they exist.
\insertfigure{loop}{\centerline{\psfig{figure=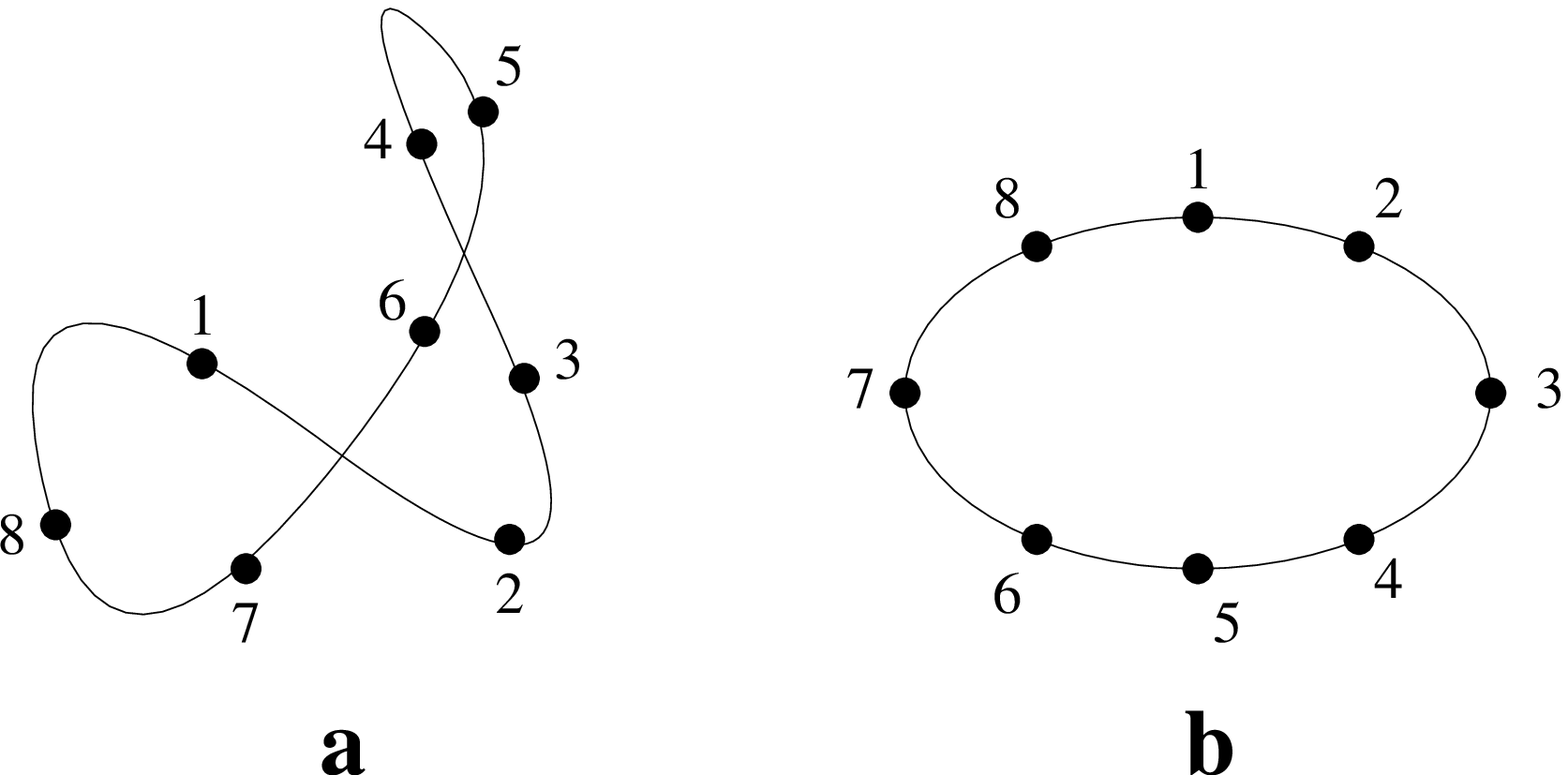,height=4cm}}}
{a) Particles ordered around a loop. b) A bound chain of particles.}{}
\noindent As a first step let us concentrate on a small
part of the loop consisting of only two particles\foot{The dynamics
of a piece of the polymer loop with any number $K$ of bits can be
precisely obtained from the second quantized theory by
applying the various singlet dynamical variables to a
non-singlet Fock state of the form
$$\ket{\Psi_\alpha^\beta}=\int\prod_{k=1}^K\big(d^{2}x_kd\theta_k\big)
[\Phi^\dagger({\bf x}_1\theta_1)
\cdots\Phi^\dagger({\bf x}_K\theta_K)]_\beta^\alpha\ket{0}
\Psi_\alpha^\beta({\bf x}_1\theta_1,\cdots,{\bf x}_K\theta_K)\;
$$
and taking the large $N_c$ limit. If such a sector showed a bound
state we could call it a piece of string.}\hskip-2pt (see \Figure{link}).
This would
correspond to a single link in the chain if the two particles were
bound.
\insertfigure{link}{\centerline{\psfig{figure=link.eps,height=1.5cm}}}
{A two particle link.}
\noindent The corresponding piece of the super-wavefunction transforms
in the adjoint representation of $U(N_c)$ and has four components,
$$\eqalign{
\Psi_\alpha^\beta(\bx_1\theta_1,\bx_2\theta_2) =
    u_1(\bx_1,\bx_2)
  &+ (\theta_1+\theta_2)u_2(\bx_1,\bx_2)\cr
  &+ (\theta_1-\theta_2)u_3(\bx_1,\bx_2)
  + \theta_1\theta_2u_4(\bx_1,\bx_2) \; ,\cr}
$$
corresponding respectively to the boson-boson, two boson-fermion,
and fermion-fermion wave functions.
The matrix indices have been dropped from the component wave functions
since the dynamics are $U(N_c)$ invariant and will not affect them.
The part of the Hamiltonian relevant for a single link in
the chain is given by
$$H_{\rm link} = -{\nabla^2\over m}
   + {1\over mr^2}\Big[2i\alpha_1\br\cdot\nabla
         - 2i\alpha_2\br\times\nabla + |\alpha|^2\Big]
  + {2\pi i\over m}(\alpha_1\pm i\alpha_2)\delta^{(2)}(\br) \; ,
\(hamlink)
$$
where $\br=\bx_1-\bx_2$ and $\nabla=(\nabla_1-\nabla_2)/2$.
The upper sign in the coefficient of the $\delta$-function holds
for $u_1$ and $u_2$, and the lower sign holds for $u_3$ and $u_4$.
Note that the two terms in the Hamiltonian
proportional to $\alpha_1$ are not separately hermitian, but
their sum is, due to the identity
$$\nabla\cdot{\br\over r^2} = 2\pi\delta^{(2)}(\br) \; .\(deltaid)$$

The above Hamiltonian contains no dimensionful parameters (other than
the mass), and therefore implies classically scale-invariant dynamics.
It appears therefore that a bound state of
{\bf finite} energy is precluded.
However we know from the simpler problem of the $\delta$-function
potential\[thornweeparton], that regularization of the contact
interaction and an interpretation of the coupling constant as a
bare parameter which depends on the regulator can yield a bound
state of finite energy depending on the regulator. To analyze the
problem at hand further requires a similar regularization.
We would like to choose a regularization that makes the
Schr\"odinger equation simplest to analyze.
One
such regularization is to replace the $\delta$-function at the
origin by a $\delta$-function at radius $R$\[Hagen],
$$\delta^{(2)}(\br)\rightarrow {1\over 2\pi R}\delta(r - R) \; ,$$
so that in the limit $R\rightarrow 0$ they are equal.
To ensure hermiticity of the regularized Hamiltonian we also
need to regularize the $\br\cdot\nabla/r^2$ term.
To do so we make the following replacement:
$${1\over r^2}\rightarrow{\theta(r - R)\over r^2} \; ,$$
where $\theta(r-R)$ is the step function. We choose to make
this replacement for {\bf all} the terms, since it will
simplify the analysis considerably.
The regularized Hamiltonian in radial coordinates is then
given by
$$\eqalign{
H_{\rm link} = &-{1\over m}
\bigg\{{\partial^2\over\partial r^2}
+ \big[1-2i\alpha_1\theta(r-R)\big]{1\over r}{\partial\over\partial r}
\cr
& + {1\over r^2}\Big[\Big({\partial\over\partial\varphi}
  +i\theta(r-R)\alpha_2\Big)^2 - \theta(r-R)\alpha_1^2\Big]
- {i(\alpha_1\pm i\alpha_2)\over R}\delta(r-R)\bigg\}\; .\cr}
\(regham)$$

The regularization we
propose corresponds to regularizing ${\bf V}(\br)$ by replacing
it with $\theta(r-R){\bf V}(\br)$. Since this can be done already
at the level of the supercharges \(firstquanmatrix), the above
regularized Hamiltonian is given by an anti-commutator of a
regularized supercharge and
its conjugate, and is therefore a positive definite operator.
Thus a negative energy bound state should not exist. A positive
energy bound state is not possible since the potential
vanishes at infinity.
To see this explicitly we solve the Schr\"odinger equation:
$$\eqalign{
\left[{\partial^2\over\partial r^2}+{1\over r}{\partial\over\partial r}
       +{1\over r^2}{\partial^2\over\partial\varphi^2}
       + k^2\right]u_n(r,\varphi) &=0\qquad r<R \cr
\left[{\partial^2\over\partial r^2}
       +\big(1-2i\alpha_1\big){1\over r}{\partial\over\partial r}
       +{1\over r^2}\Big[
         \Big({\partial\over\partial\varphi}+i\alpha_2\Big)^2
         - \alpha_1^2\Big]
         +k^2\right]u_n(r,\varphi) &= 0
        \qquad r>R \; .\cr}
\(Schroding)$$
The parameter $\alpha_1$ can be eliminated from the second equation
by redefining the outer wave function, resulting in the
following equations for the inner an outer
wave functions
$$\eqalign{
\left[{\partial^2\over\partial r^2}+{1\over r}{\partial\over\partial r}
       +{1\over r^2}{\partial^2\over\partial\varphi^2}
       + k^2\right]u_n(r,\varphi) &= 0 \qquad r<R \cr
\left[{\partial^2\over\partial r^2}
     +{1\over r}{\partial\over\partial r}
     +{1\over r^2}
     \Big({\partial\over\partial\varphi}+i\alpha_2\Big)^2
    + k^2\right]r^{-i\alpha_1}u_n(r,\varphi) &= 0
        \qquad r>R \; .\cr}
\(Schro)$$
Defining the radial wave functions by
$$u_n(r,\varphi)=e^{il\varphi}\chi_n(r) \; ,$$
with $l$ an arbitrary integer labeling the angular momentum,
the jump condition on the logarithmic derivatives imposed by the
$\delta$-function is given by:
$$
{\chi_n^\prime(R+\epsilon)\over\chi_n(R)}
- {\chi_n^\prime(R-\epsilon)\over\chi_n(R)} =
   {i(\alpha_1\pm i\alpha_2)\over R}\; .
\(jump)$$
For negative energy solutions we define $B\equiv -k^2>0$.
Regularity at the origin and normalizability implies the
following form for the
negative energy solution:
$$\chi_n(r)=\cases{AI_{|l|}(\sqrt Br)
                 \qquad {\rm for} \; r<R \cr
  Cr^{i\alpha_1}K_{|l+\alpha_2|}(\sqrt Br)
                 \qquad {\rm for} \; r>R \; ,\cr}
\(regularsol)$$
where the constants $A$ and $C$ are determined by the continuity
condition at $r=R$ and by normalization.
The jump condition
\(jump) then gives the following:
$${K_{|l+\alpha_2|}^\prime(\sqrt BR)\over K_{|l+\alpha_2|}(\sqrt BR)}
 -{I_{|l|}^\prime(\sqrt BR)\over I_{|l|}(\sqrt BR)} =
 \mp{\alpha_2\over\sqrt BR} \; .
\(relation)$$
Using the following recurrence relations for the modified
Bessel functions,
$$\eqalign{
K_\nu^\prime(z) &= -K_{\nu -1}(z) - {\nu\over z}K_\nu(z) \cr
I_\nu^\prime(z) &= I_{\nu +1}(z) + {\nu\over z}I_\nu(z) \; ,\cr}
\(recurrence)$$
gives the condition
$$\sqrt BR\left[
{K_{|l+\alpha_2|-1}(\sqrt BR)\over K_{|l+\alpha_2|}(\sqrt BR)}
 +{I_{|l|+1}(\sqrt BR)\over I_{|l|}(\sqrt BR)}\right] =
 \pm\alpha_2-|l|-|l+\alpha_2| \; .
\(relation2)$$
Since the functions $K_\nu(z)$ and $I_\nu(z)$ with $\nu>-1$
are positive for
$z>0$ the left hand side of the
equation is positive for $\sqrt BR>0$.
When $\sqrt BR = 0$ the left hand side vanishes.
The right hand side is clearly negative or zero, since
$$\pm\alpha_2 - |l| \leq |l+\alpha_2| \qquad {\rm for\;\; all\;\; }l \; .$$
Consequently there is no solution except when the above inequality
is saturated, in which case the bound state energy vanishes.
Since the two particles comprising a link in the chain cannot
bind, a closed chain will not form. This is therefore, as we
expected, an unsatisfactory model for describing discretized
superstring.
\chapter{Discussion}
We have presented field theoretic representations of the full
($\sgtwo$) Super-Galilei algebra of space-time transformations,
both with single component and matrix valued fields.
In the first case we showed that non-relativistic
Super-Chern-Simons theory emerges as a special case of our model.
The matrix field
theory is motivated by the discretized light-cone superstring, but
fails to be a satisfactory model since closed polymer chains, which
become strings in the continuum limit, do not form.
The two-body Hamiltonian is positive definite, which precludes any
negative energy bound states. For zero energy or positive energy bound states
to exist, we must have a potential energy which is positive and
non-vanishing at infinite separation.
In a separate
paper we present an $\sgone$ invariant matrix field theory
which achieves a satisfactory free superstring limit, because
it employs a harmonic potential between string bits. Although
the large $N_c$ limit of that model had the full $\sgtwo$
invariance, the symmetry was broken to $\sgone$ at finite $N_c$,
and $\sgone$ invariance is not sufficient
to force the correct superstring interactions. Thus the ultimate
goal should be
to build a satisfactory string bit theory with the full $\sgtwo$
Super-Galilei symmetry at all values of $N_c$.
\subsection{Acknowledgments:} We should like to thank
Kostas Anagnostopoulos, Zongan Qiu, and Pierre Ramond
 for useful and enlightening discussions.

\bigskip
\titlestyle{References}
\reflist{}
\end